\documentclass[preprint,12pt]{elsarticle}

\usepackage{amssymb}

\journal{Physica A: Statistical Mechanics and its Applications}

\usepackage[hyphens]{url} % split url
\usepackage{subfig} % \subfloat
\usepackage{rotating} % Rotating table with sidewaystable
\usepackage{comment}
\usepackage{tikz}
\usepackage{float}
\usepackage[hyphens]{url} % split url
\usepackage{subfig} % \subfloat

\usepackage{multicol}
\usepackage[utf8]{inputenc}
\usepackage{csquotes}

\usepackage{caption}
\usepackage{amsmath}
\usepackage{changepage}
\usepackage{courier}
\usepackage{colortbl} % \arrayrulecolor{gray} hline color

\usepackage{booktabs} % Required for better horizontal rules in tables

\usepackage{float}
\usepackage{placeins} %\FloatBarrier
\usepackage{geometry} % \newgeometry \restoregeometry
\usepackage{verbatim}
\usepackage{xcolor} %define color
\definecolor{light-gray}{gray}{0.92}
\usepackage{listings} %R code
\usepackage{listingsutf8}
\usepackage{booktabs} %/tourples

\usepackage{tabularx}

\usepackage{graphicx}
\usepackage{pdflscape}
\usepackage{adjustbox}

\usepackage[labelfont=bf]{caption} % titulo Figure aparece en negritas

\usepackage[labelsep=period]{caption}
\usepackage[noabbrev]{cleveref} 			% para melhorias de justificação

\begin{document}

\begin{frontmatter}



	% SUGESTAO PARA O TITULO:
    \title{Optimal rewiring in coupled opinion and epidemic dynamics with vaccination}


	 \author[eco]{André L. Oestereich\corref{cor1}}

    \author[UFAL]{Marcelo A. Pires}

	 \author[uff]{Nuno Crokidakis}

	 \author[eco,INCT,LAMFO]{Daniel O. Cajueiro}

	 \address[eco]{Department of Economics, University of Bras\'ilia, Campus Universitario Darcy Ribeiro - FACE, Bras\'ilia, DF 70910-900.}

     \address[UFAL]{Universidade Federal de Alagoas, 57480-000, Delmiro Gouveia - AL, Brasil}

	 \address[uff]{Instituto de F\'isica, Universidade Federal Fluminense, Niter\'oi/RJ, Brazil}

	 \address[INCT]{National Institute of Science and Technology for Complex Systems (INCT-SC), Brazil.}

	 \address[LAMFO]{Machine Learning Laboratory in Finance and Organizations (LAMFO), Universidade de Bras\'{i}lia (UnB), 70910-900, Bras\'{i}lia, Brazil.}

	 %\author{Daniel Oliveira Cajueiro\corref{cor1}}

	 \cortext[cor1]{Corresponding author: andrelo@id.uff.br }

	\begin{abstract}
        In this work, we study an epidemic model with vaccination coupled with opinion dynamics in a dynamic network. The network structure evolves as agents with differing opinions disconnect from one another and connect with agents that share similar opinions about vaccination. We consider a SIS-like model with an extra vaccinated state. Agents can have continuous opinions and every time an agent disconnects from a neighbor, they connect to a new neighbor. We have observed the emergence of network homophily and, in certain cases, the complete fragmentation of the network. Our Monte Carlo simulations also show first-order phase transitions with metastable states.      An increase in the probability of rewiring yields a dual effect, namely: (a) in the short term, it has the potential to intensify the epidemic peak; (b) in the long term, it can diminish the rate of infection. This transient increase in the epidemic peak is attributed to the fragmentation of the network into smaller, disconnected sub-networks. Therefore, our results suggest that rewiring is optimal when it is as high as possible but before the network starts breaking apart.
	\end{abstract}

\begin{highlights}
\item Existence of an optimal rewiring
\item Bistability and metastable states
\item Nonmonotonic effects
\end{highlights}


\end{frontmatter}

\section{Introduction}

Statistical physics is a scientific field that typically addresses macroscopic phenomena arising from the microscopic interactions among constituent units.
Researchers have shown over the years that the framework of statistical physics provides a valuable toolbox for describing phenomena that go beyond the traditional domain of physics~\cite{2009castellanoFL,2016wangBBD}.

Within the framework of collective phenomena, a complex problem arises: what are the conceivable macroscopic scenarios that can result from the intricate interplay between vaccination, epidemics, and opinion dynamics? This topic is of great significance as the effectiveness of a vaccination campaign is not solely determined by factors such as vaccine accessibility, efficacy, and epidemiological variables but also depends on public opinion regarding vaccination~\cite{dube2015strategies}.
The integration of opinion dynamics and epidemic models enables the capture and comprehension of novel phenomena that cannot be represented by conventional models of epidemic spreading, such as the SIS, SIR, and SIRS models\cite{2015wangAWW,pires2017dynamics,2018piresOC,2021piresOCQ,10.1093/pnasnexus/pgac260,leung2022impact}.

Our study aims to go beyond previous research efforts~\cite{pires2017dynamics,2018piresOC,2021piresOCQ} by incorporating an additional dynamical process, namely an evolving network.  Specifically, each individual within the proposed model is endowed with the capacity to adapt its immediate neighborhood in accordance with the opinions of its interacting peers.

Through Monte Carlo simulations, we investigate various scenarios that depend on epidemic parameters, rewiring characteristics, and opinion dynamics attributes. Our findings demonstrate a first-order phase transition with metastable states. Besides the common processes that are modelled with epidemilogical models and opinion dynamics models, we also observe that the rewiring process promotes network homophily, which influences the short-term and long-term dynamics of epidemics. 

\section{Literature review}

We consider an agent-based computational model to study a multi-coupled dynamics that incorporates the interplay between opinion dynamics, epidemic spreading, vaccination, and rewiring dynamics. Agent-based computational models are valuable tools for dealing with complex problems that arise in society~\cite{2012squazzoni,2015bianchiS,2012conteGBC}. By simulating the behavior of a system, these models can provide a comprehensive view of possible outcomes that may arise under varying conditions and parameter values.

Opinion dynamics is a scientific field that has significantly contributed to the understanding of the process of opinion formation, evolution, and diffusion \cite{2012galam,2008galam,2014senC,2019oestereichPC,oestereich2020hysteresis,pires2022double}. Initially based on typical physical systems, opinion dynamics models evolved to include agents' roles and behaviors, such as inflexibility, independence, and contrarian behavior. These models also consider external influences, such as fake news, information manipulation, algorithmic filtering, and more%others on opinion formation and evolution
~\cite{sobkowicz2020whither}. For recent reviews on opinion dynamics modeling, see \cite{zha2020opinion,xia2011opinion,anderson2019recent,DONG201857,sirbu2017opinion,peralta2022opinion,biswas2023social}.

Epidemics and vaccination dynamics are important areas of research that aim to understand epidemic processes and assist in the control of infectious diseases~\cite{2016wangBBD}. The SIS (susceptible-infected-susceptible), SIR (Susceptible-infectious-removed), and other compartmental models, provide a paradigmatic framework for describing the spread of an epidemic. To study vaccination, within these models an additional compartment $V$ can be added to represent immunity gained through vaccination or the compartment $R$ can be modified to include vaccine-induced immunity.

Rewiring dynamics refer to the process by which individuals in a network change their connections over time. Such changes can be driven by factors such as individual behavior, external events, or evolutionary pressures~\cite{gross2008adaptive}.
When rewiring is introduced in epidemic dynamics, novel phenomena emerge. For instance, in the seminal work~\cite{2006grossDB}, researchers investigated the SIS model in a dynamic network, in which susceptible individuals can change their neighborhood by rewiring their connections to avoid contact with infected individuals.
They showed the emergence of rich phenomenology from the interplay between dynamics and structure.
In the context of the current study, rewiring dynamics plays a role in shaping the network topology, which affects the dynamics of opinion formation and epidemic spreading.

\section{Model}

The model consists of a variant of the coupled opinion with epidemic model developed recently \cite{2018piresOC}.
The main difference in this model is that now agents with disagreeing opinions can be disconnected. We considered an Erdos-Renyi random graph where each of the $N$ nodes is occupied by an individual and has $M=Nk$ connections. Each agent $i$ in this society carries an opinion $o_i$, that is a real number in the range $[-1, +1]$.
Positive (negative) values indicate that the position is favorable (unfavorable) to the vaccination campaign \cite{2018piresOC}.
Opinions tending to $\,+1$ and $\,-1$ indicate extremist individuals.
Finally, opinions near $0$ mean neutral or undecided agents.
We will consider an epidemic dynamics coupled with the opinion dynamics regarding the vaccination, with the agents being classified as follows:

\begin{itemize}
    \item Opinion states:
    Pro-vaccine (opinion $o_i > 0$) or Anti-vaccine (opinion $o_i < 0$) individuals;
    \item  Epidemic compartments: Susceptible (S), Infected (I) or Vaccinated (V) individuals;
\end{itemize}

After the introduction to the basic composition of the model, the algorithm that governs the dynamics of the model is as follows:

\begin{enumerate}
    \item Initial State: set a fraction $D$ of the opinions as +1 and the rest as -1; set 10\% of agents as INFECTED and the rest as SUSCEPTIBLE.
    \item Opinion update:
    \begin{enumerate}
        \item select a random agent $i$ and one of its neighbors $j$;
        \item if $|o_i - o_j| > \epsilon$ and with probability $r$ if possible the connection between $i$ and $j$ is removed and $i$ is connected to a valid random agent $k$ such that $|o_i - o_k| < \epsilon$. If a new connection is impossible we keep the previous connection in order to preserve the average connectivity of the network;
        \item otherwise update agent's opinion to $o_i = o_i + \eta o_j + w I_i$, where $I_i$ is the infection rate of the neighbors of $i$ (i.e., $I_i$ is the proportion of Infected neighbors of agent $i$), $w$ is the risk perception parameter and $\eta$ is a random variable in the range $[0,1]$. So, in case the agent is not rewired they interact with their chosen neighbor.
    \end{enumerate}
    \item  Epidemic update:
    \begin{enumerate}
        \item select random agent $i$;
        \item if agent $i$ is SUSCEPTIBLE
            \begin{enumerate}
                \item they become VACCINATED with probability $(o_i+1)/2$;
                \item otherwise agent $i$ becomes INFECTED with probability $\lambda I_i$. Since $I_i$ is the local infection rate, the infection probability depends on the agents to which agent $i$ is connected to;
            \end{enumerate}
        \item if agent $i$ is INFECTED they become SUSCEPTIBLE with probability $\alpha$;
        \item if agent $i$ is VACCINATED they become SUSCEPTIBLE with probability $\phi$.
    \end{enumerate}
    \item Repeat the previous two steps $t$ times, where $t$ is the number of Monte Carlo steps.
\end{enumerate}

A summary of the model's parameters can be seen in \cref{Tab1} as well as their meaning and default values. The results presented in the next section were obtained from the average of multiple Monte Carlo simulations.

\begin{table*}[tbp]
\begin{center}
    \vspace{0.1cm}
    \renewcommand\arraystretch{1.3}
    \begin{tabular}{|c|c|c|}
        \hline \textbf{Variable} & \textbf{Default value} & \textbf{Explanation } \\
        \hline $N$ & 10000 & number of agents \\
        \hline $k$ & 50 & average number of connections per agent \\
        \hline $D$ &  & fraction of initial opinions +1 \\
        \hline $r$ & & probability of rewiring \\
        \hline $\epsilon$ & 1 & distance between opinions to be rewired \\
        \hline $w$ &  & intensity of risk perception \\
        \hline $\lambda$ &  & infection rate \\
        \hline $\alpha$ & 0.1  & recovery rate \\
        \hline $\phi$ & 0.01  & vaccine waning rate \\
        \hline
    \end{tabular}
\end{center}
\caption{Model's parameters and the meaning of each one. From now the parameters are set to their default values unless explicitly mentioned.}
\label{Tab1}
\end{table*}

\section{Results}

\begin{figure}[h!]
    \begin{center}
        \subfloat[$r=0.4$]{
            \includegraphics[width=0.34\textwidth]{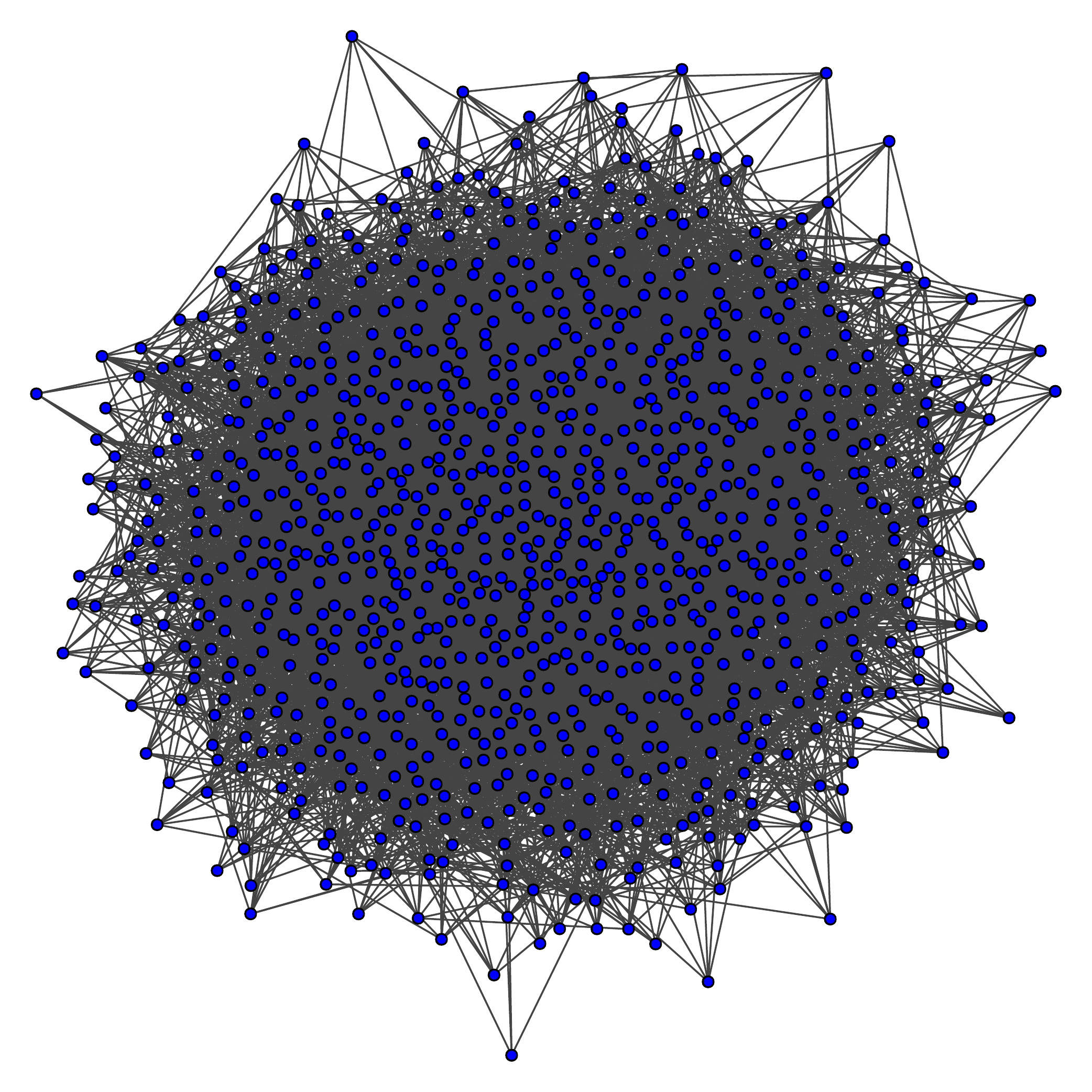}
        }
        \subfloat[$r=0.6$]{
            \includegraphics[width=0.34\textwidth]{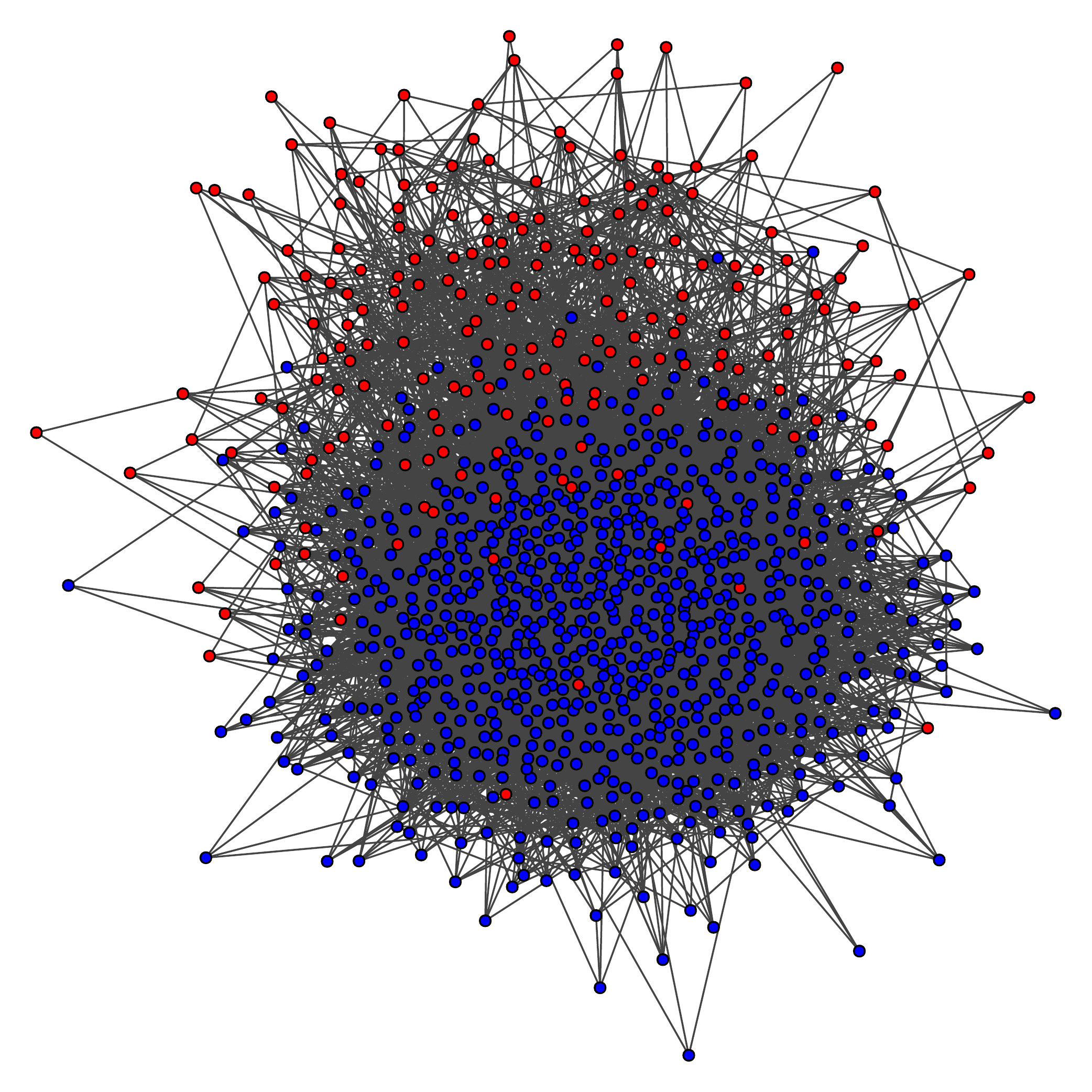}
        } \\
        \subfloat[$r=0.8$]{
            \includegraphics[width=0.34\textwidth]{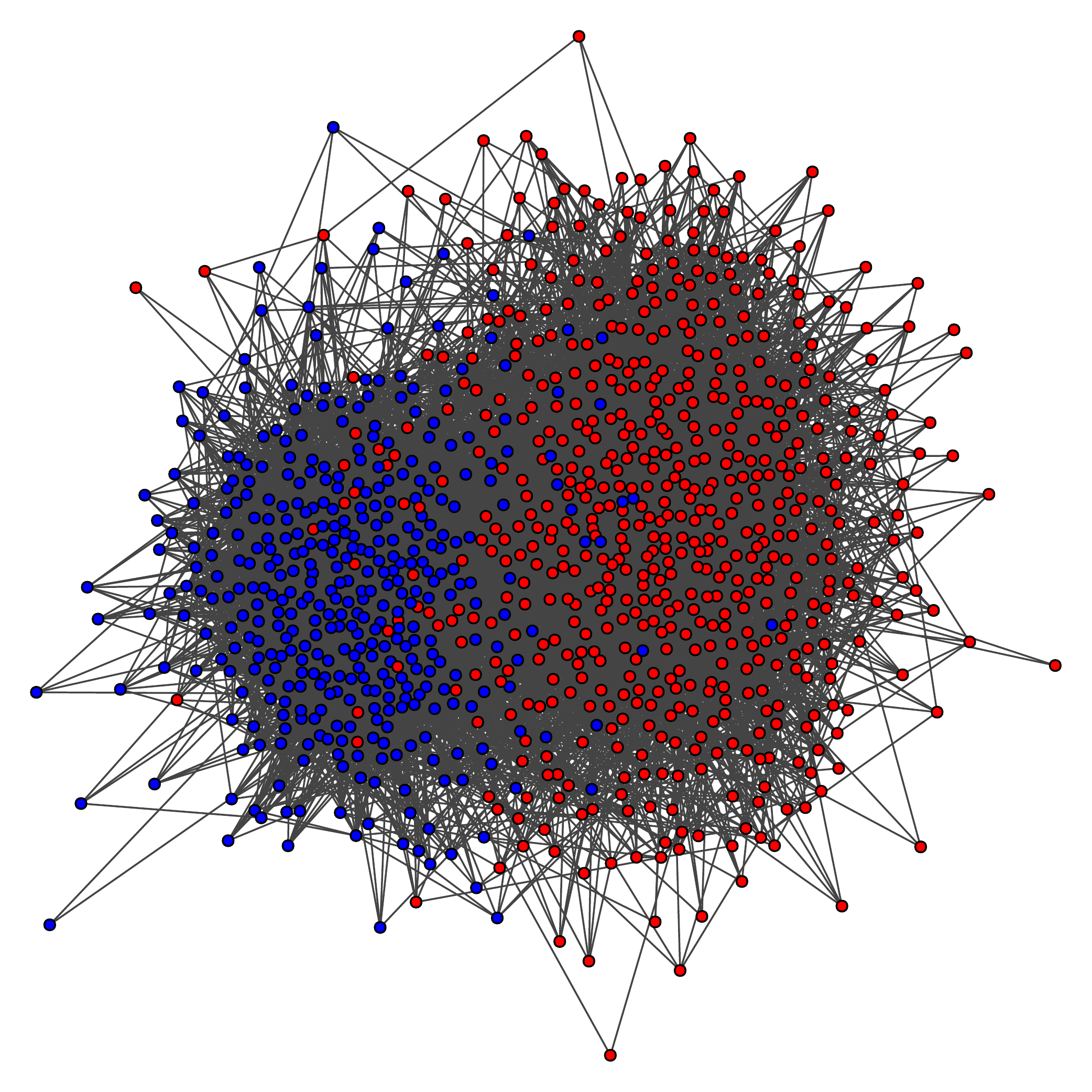}
        }
        \subfloat[$r=1.0$]{
            \includegraphics[width=0.34\textwidth]{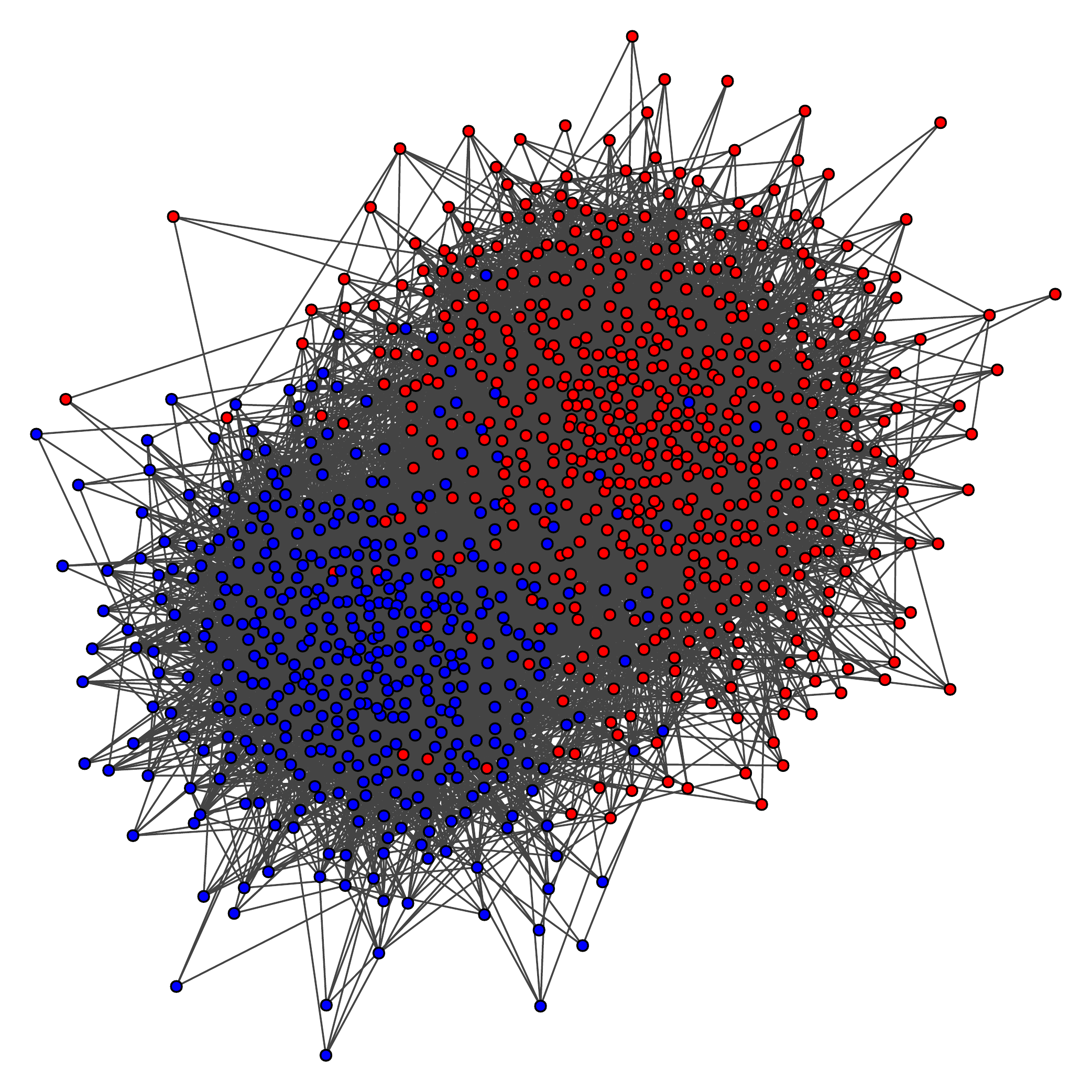}
        }
    \end{center}
    \caption{Graphs of the opinion distribution in the stationary state for $N=1000, k = 5, w=0.4, \lambda=0.4$ and $D=0.4$ where the red dots are anti-vaccine individuals and blue dots are pro-vaccine individuals.}
    \label{graphs}
\end{figure}

In \cref{graphs}, we can see examples of the stationary state of graphs with $N=1000$ formed by the opinion dynamics. It is noticeable that this dynamic tends to separate the distinct opinions into communities that can become disconnected from the rest of the nodes. As the rewiring parameter ($r$) increases we also see an increase in the homophily of the network. In more extreme cases, this can lead to the fragmentation of the network into several small networks where all nodes share the same opinion. With increasing homophily in the network achieving consensus becomes harder, so increases in $r$ tend to result in an average opinion closer to zero. This is corroborated by results presented in subsequent sections.

    %These can have various sizes but as $r$ increases they tent to be of similar size. The size between the communities can also be inferred from the average opinion of the system, since similar sizes tend to bring the average closer to zero.

\begin{figure}[h]
    \centering
    \includegraphics[width=0.44\textwidth]{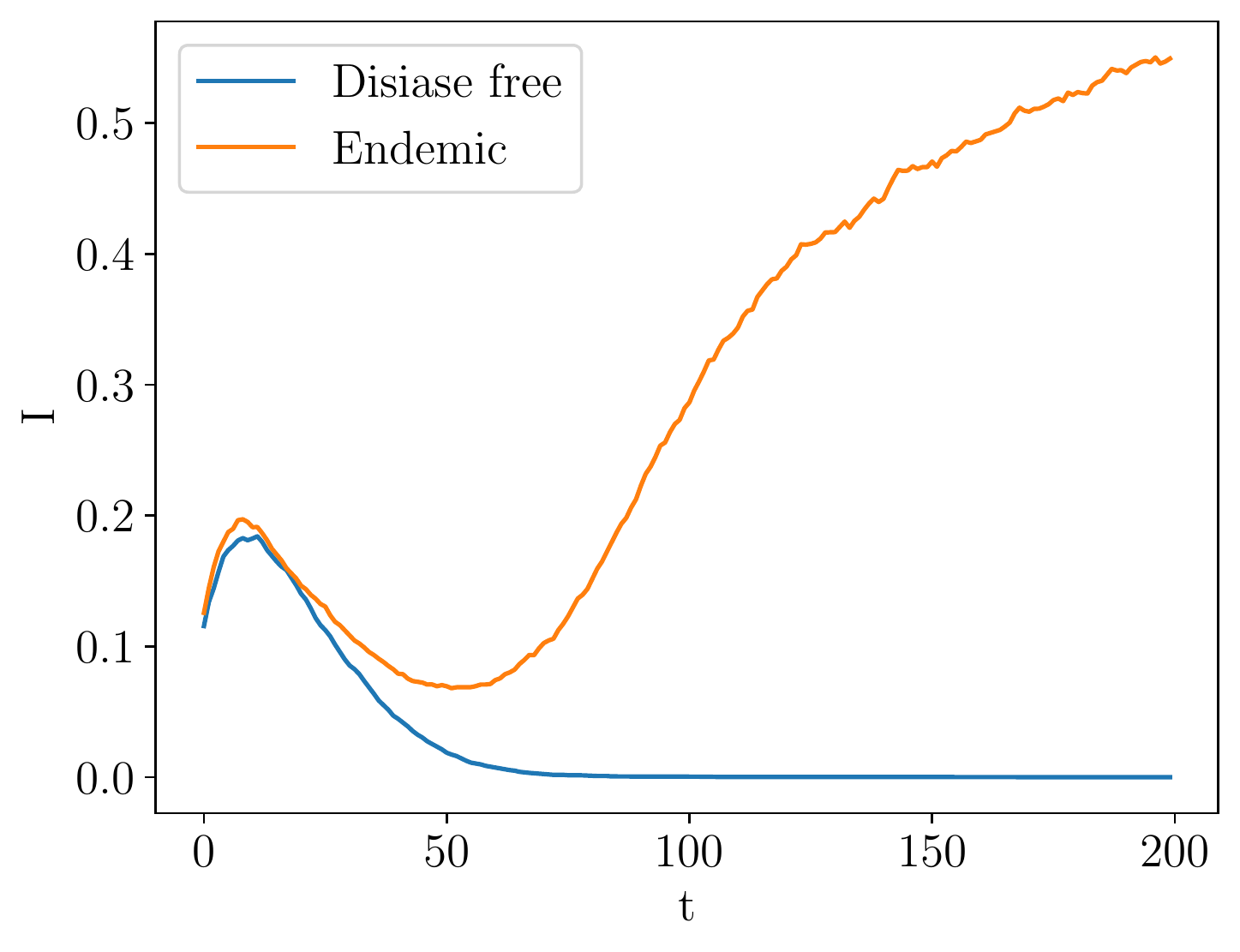}
    \includegraphics[width=0.44\textwidth]{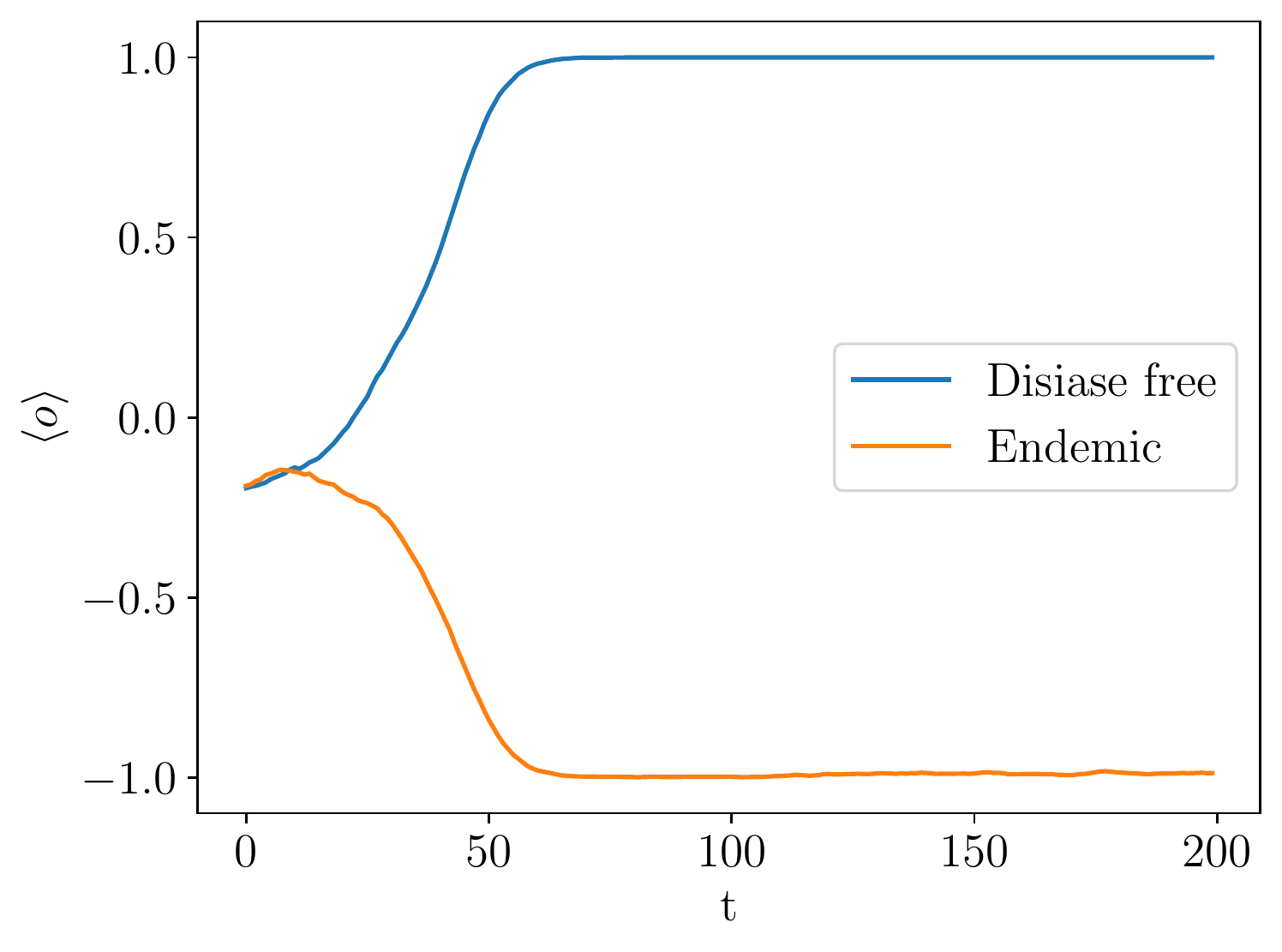}
    \caption{Time series for two samples, one in which the disease spreading is stopped by the vaccination due to positive opinion about it and one in which the disease persists due to negative opinion about the vaccination. Time is measured in Monte Carlo steps and both samples have $w=0.3$, $D=0.4$, $r=0.5$ and $\lambda = 0.5$.}
    \label{Ixt}
\end{figure}

Evidence of meta-stability can be observed in \cref{Ixt}, where we present the time series of two samples with the same configuration, one of which ended in the endemic state, and another that ended disease-free. In this figure, we can also see that for this parameter configuration the defining factor in the direction of the dynamic is the opinion, as its separation precedes the separation in the average infection rate. This is the case because the value for the risk perception ($w$) is relatively low, so the feedback from the infection to the opinions is not too strong.

The same general pattern generally repeats in the time series. Initially, without any vaccinated agents in the first few time steps, $I$ increases. After the initial vaccination of the pro-vaccine agents, $I$ decreases unless the initial pro-vaccine fraction of agents is too small. In the final stage, the result depends on the stationary average opinion of the agents.

In the next subsections, we consider distinct values of $D$, the initial fraction of positive (pro-vaccine) opinions. Similarly to what happens in \cite{2018piresOC}, the opinion dynamics for cases with $D > 0.5$ leads to a consensus in $o_i = 1 \forall i$ since the overall opinion is positive (pro-vaccine), $\eta\geq 0$, and $w I_i \geq 0$. This implies that every agent will certainly vaccinate, which, in turn, stops the epidemic spreading. Thus, we will consider the more interesting scenarios in which the initial majority is against vaccination, i.e., $D < 0.5$, to isolate and better understand the effects of the other variables.

\FloatBarrier
\subsection{$D = 0.0$}

\begin{figure}[h]
    \centering
    \subfloat[Average opinion]{
        \includegraphics[width=0.44\textwidth]{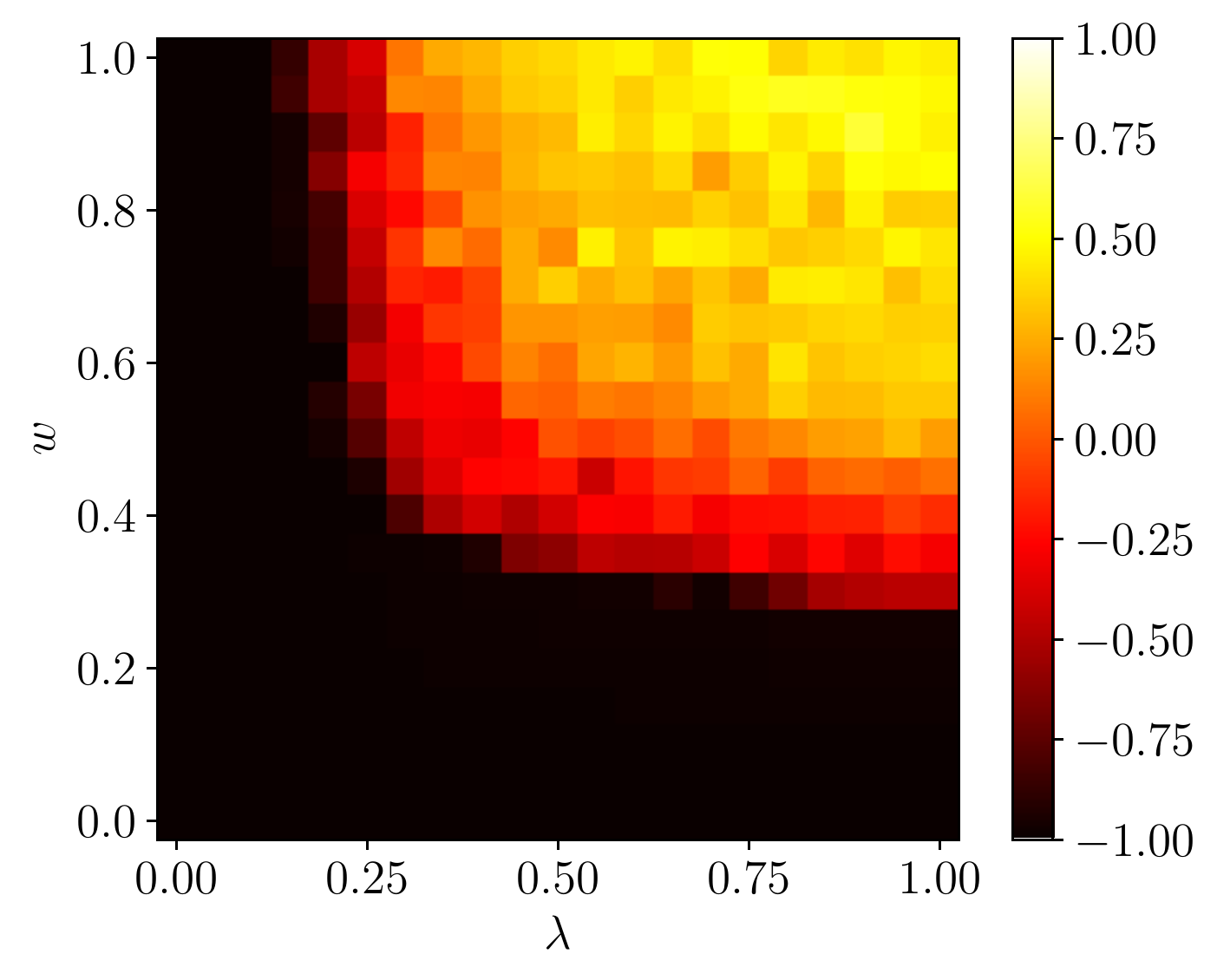}
    }
    \subfloat[Infected rate]{
        \includegraphics[width=0.44\textwidth]{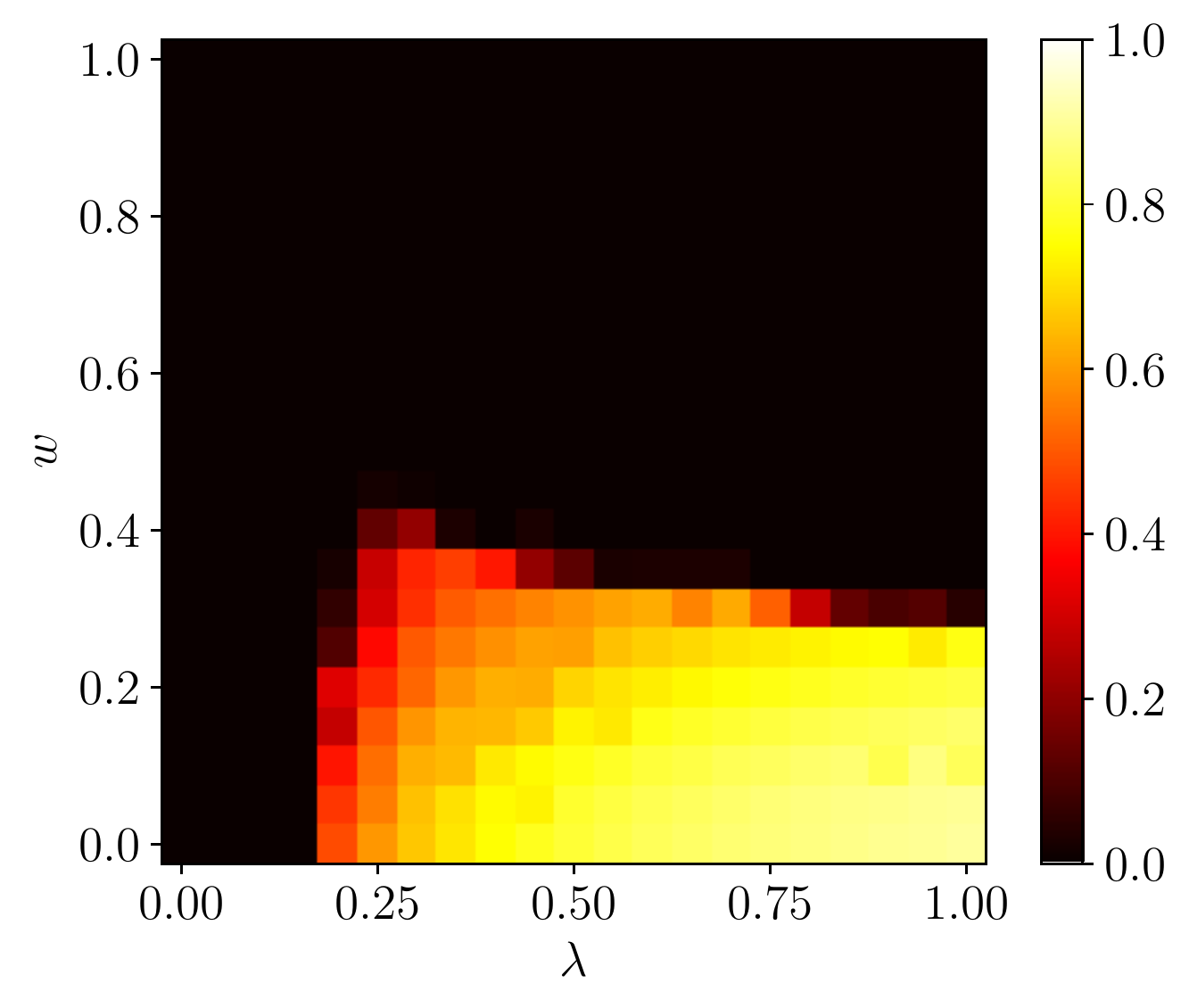}
    }
    \caption{Average opinion and infected rate vs infection rate and risk perception, for $r=1$ and $D=0$. In this setting all agents start with the same opinion and therefore the rewiring does not have an appreciable effect.}
    \label{LxW_D2}
\end{figure}

$D$ is the parameter that sets the initial disagreement between agents. Therefore, for $D=0.0$, we have unanimous anti-vaccine extremism, $o_{i} = -1$, $\forall i$. In this case, the rewiring parameter is not especially relevant, since agents are in consensus, and rewiring can not occur regardless of the rewiring parameter. The results in this case are very similar to those presented in \cite{2018piresOC}, and can be seen in \cref{LxW_D2}.

Regardless of the rewiring probability, there must be a minimum distance $\epsilon$ between the opinions of the agents. This means that when the opinions are grouped too closely together, rewiring is not possible. Therefore, in these case we end up with a networked version of the previous model \cite{2018piresOC}, with similar results.

This grouping of opinions is somewhat expected, since the LCCC model presents low variance of the opinions in the stationary state \cite{2010lallouacheCCC}, and the external field acts somewhat uniformly in each region of the graph. As the rewiring model depends on a certain distance between opinions, this distance must be established in the initial state, and resulting in a strong dependence with the initial state.

\FloatBarrier
\subsection{$D=0.2$}

With $D=0.2$, we now have distinct starting opinions, and rewiring can occur. As we still have a clear initial majority, rewiring tends to manifest mostly for higher values of $r$. The rewiring starts causing a segmentation between agents with different opinions. This, in turn, leads to the preservation of both opinions in the system and, therefore, a more moderate average opinion.

\begin{figure}[h]
    \centering
    \subfloat[$r=0.5$]{
        \includegraphics[width=0.32\textwidth]{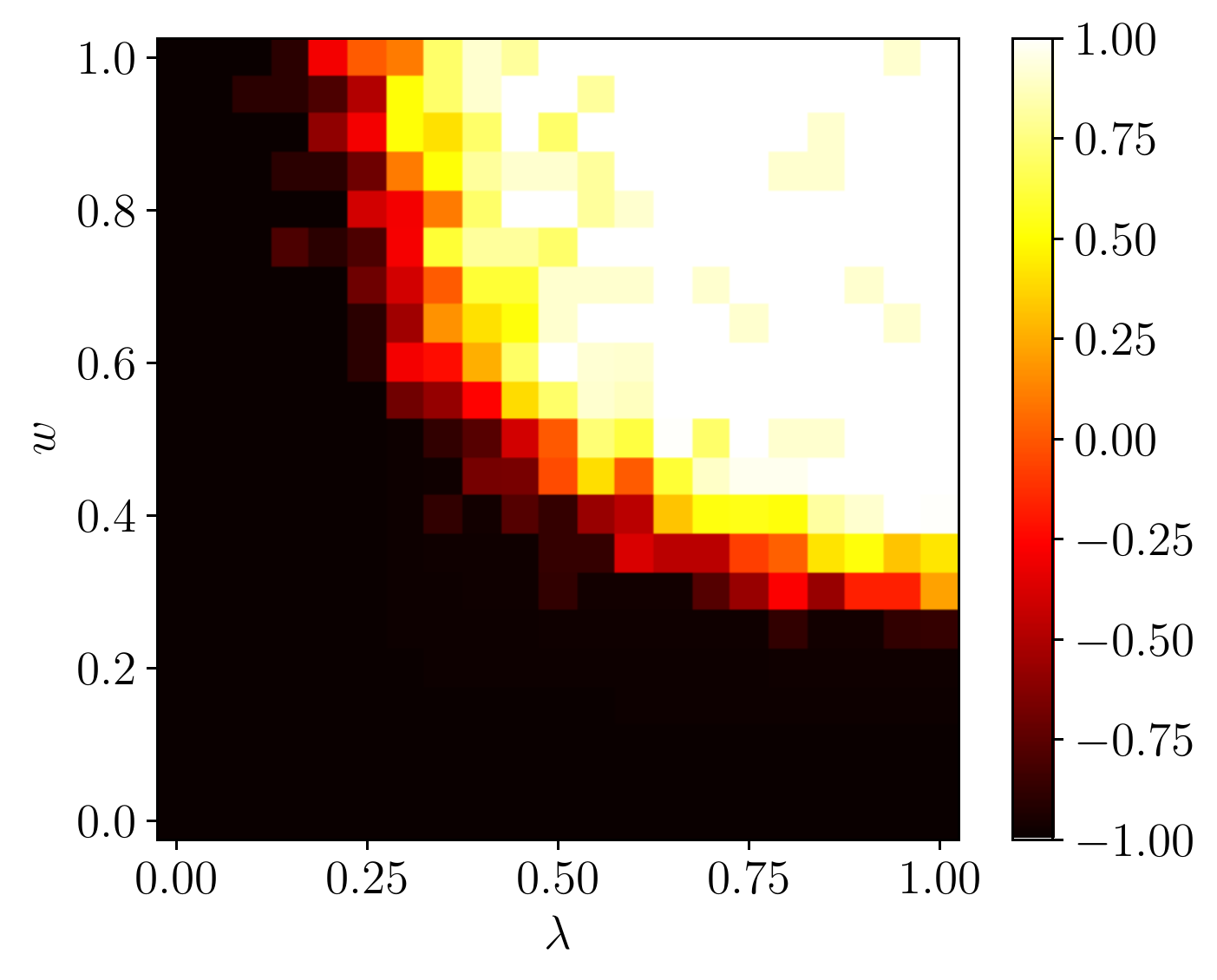}
    }
    \subfloat[$r=0.6$]{
        \includegraphics[width=0.32\textwidth]{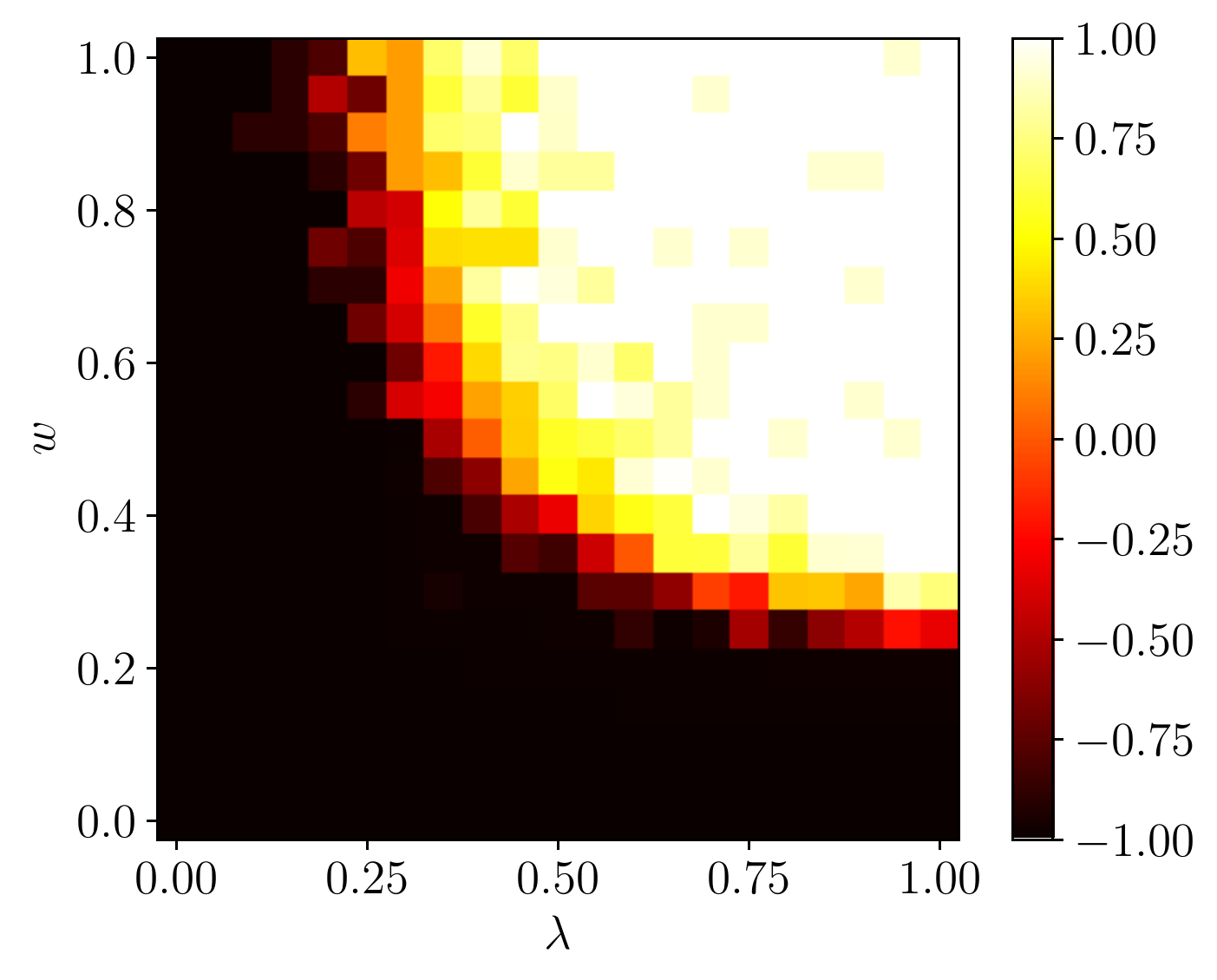}
    }
    \subfloat[$r=0.7$]{
        \includegraphics[width=0.32\textwidth]{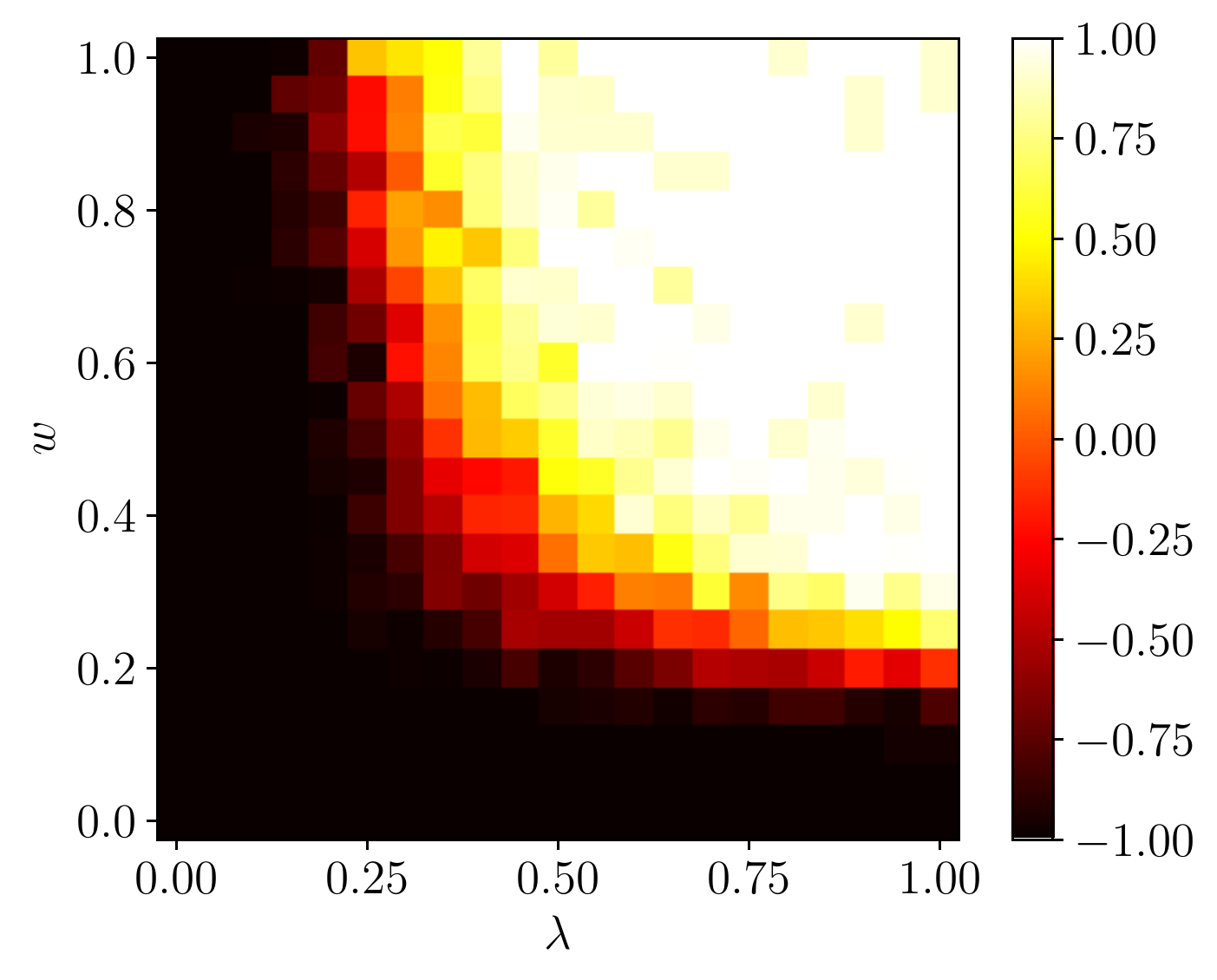}
    } \\
    \subfloat[$r=0.8$]{
        \includegraphics[width=0.32\textwidth]{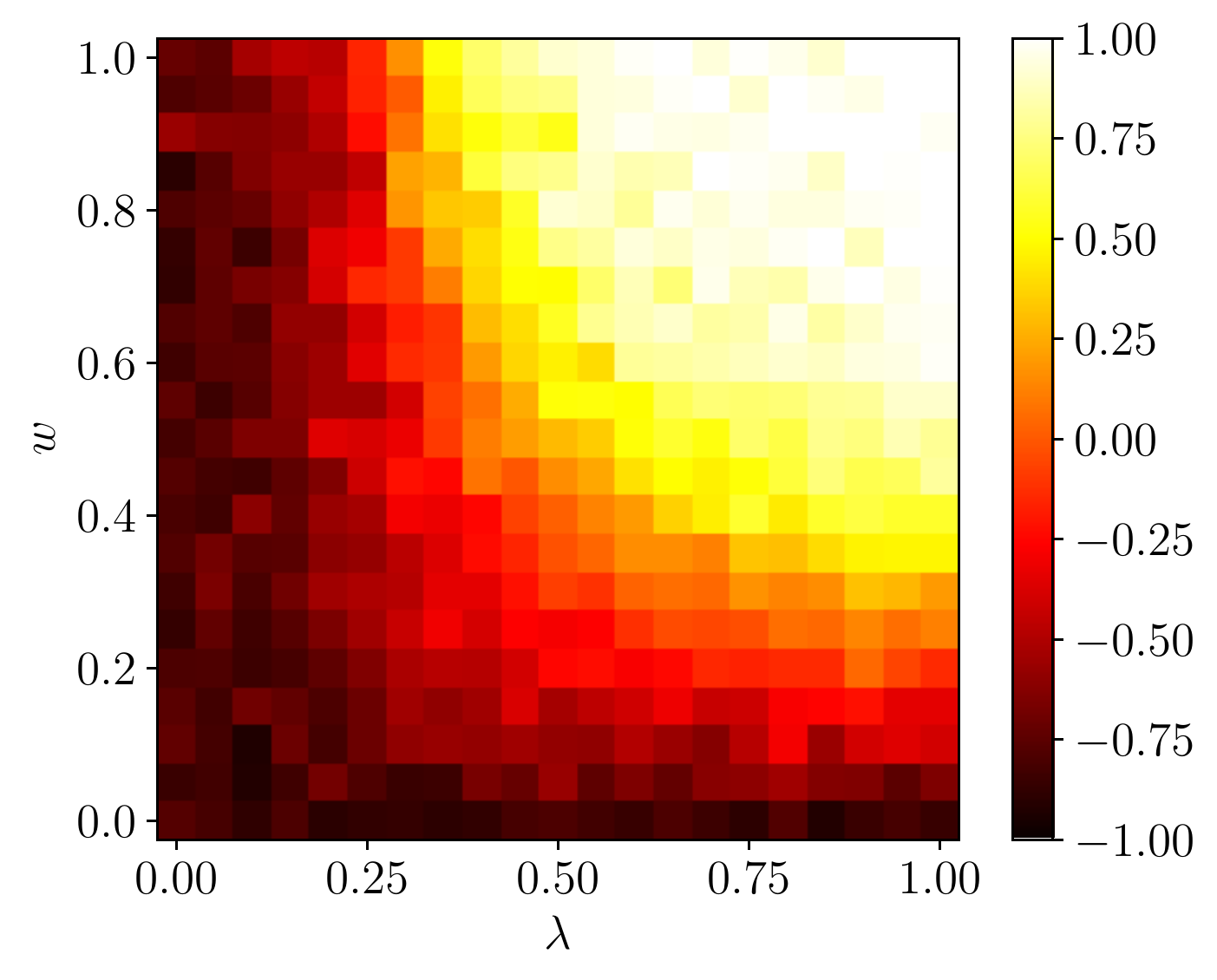}
    }
    \subfloat[$r=0.9$]{
        \includegraphics[width=0.32\textwidth]{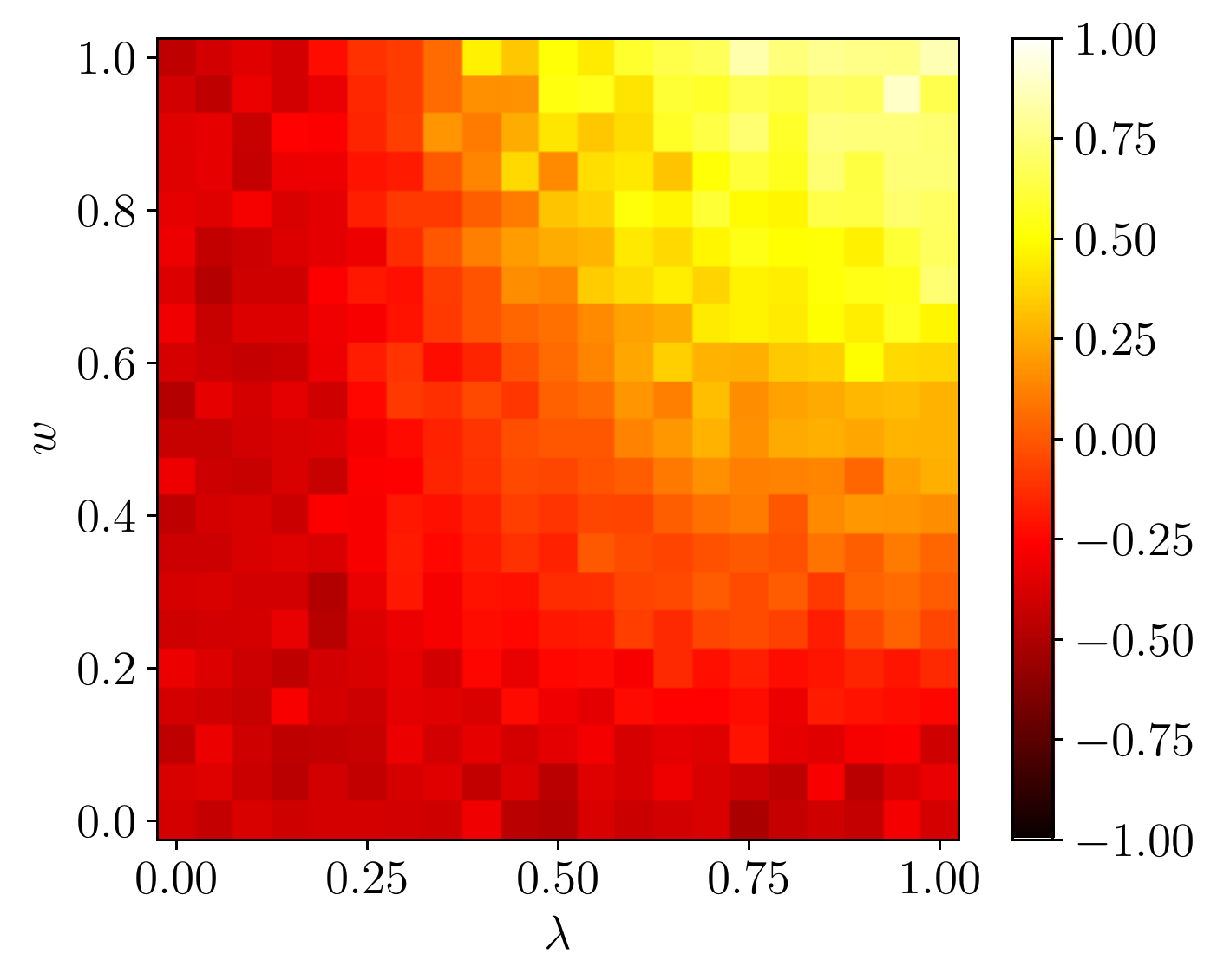}
    }
    \subfloat[$r=1.0$]{
        \includegraphics[width=0.32\textwidth]{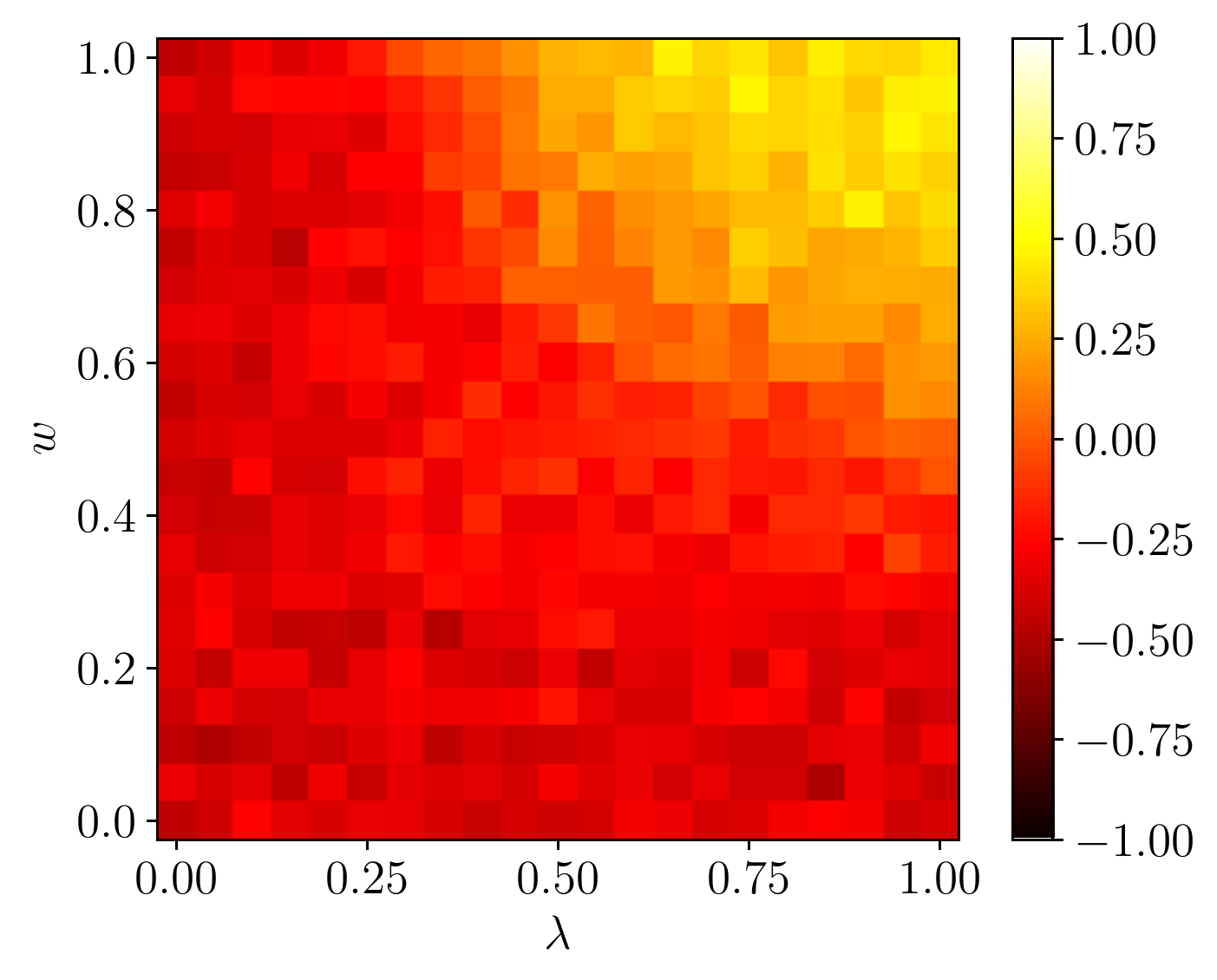}
    }
    \caption{Average opinion vs infection rate and risk perception for $D=0.2$ and various rewiring probabilities. With some amount of difference between the initial opinions the rewiring probability starts to have a noticeable effect in the stationary state, i.e., for higher values of $r$ consensus either pro or against vaccination becomes less likely.}
    \label{OxWxL_D2}
\end{figure}

\Cref{OxWxL_D2} shows the average opinion in the stationary state. In this figure we notice that as $r$ increases the region of moderate opinion starts to grow. Firstly in the surroundings of the phase transition and then for $r \approx 1$ it takes over the whole region. This is where the double-edged nature of this strategy lies. In one hand the rewiring reduces the chances for anti-vaccine consensus, but in the other it also reduces pro-vaccine consensus.

\begin{figure}[h]
    \centering
    \subfloat[$r=0.5$]{
        \includegraphics[width=0.32\textwidth]{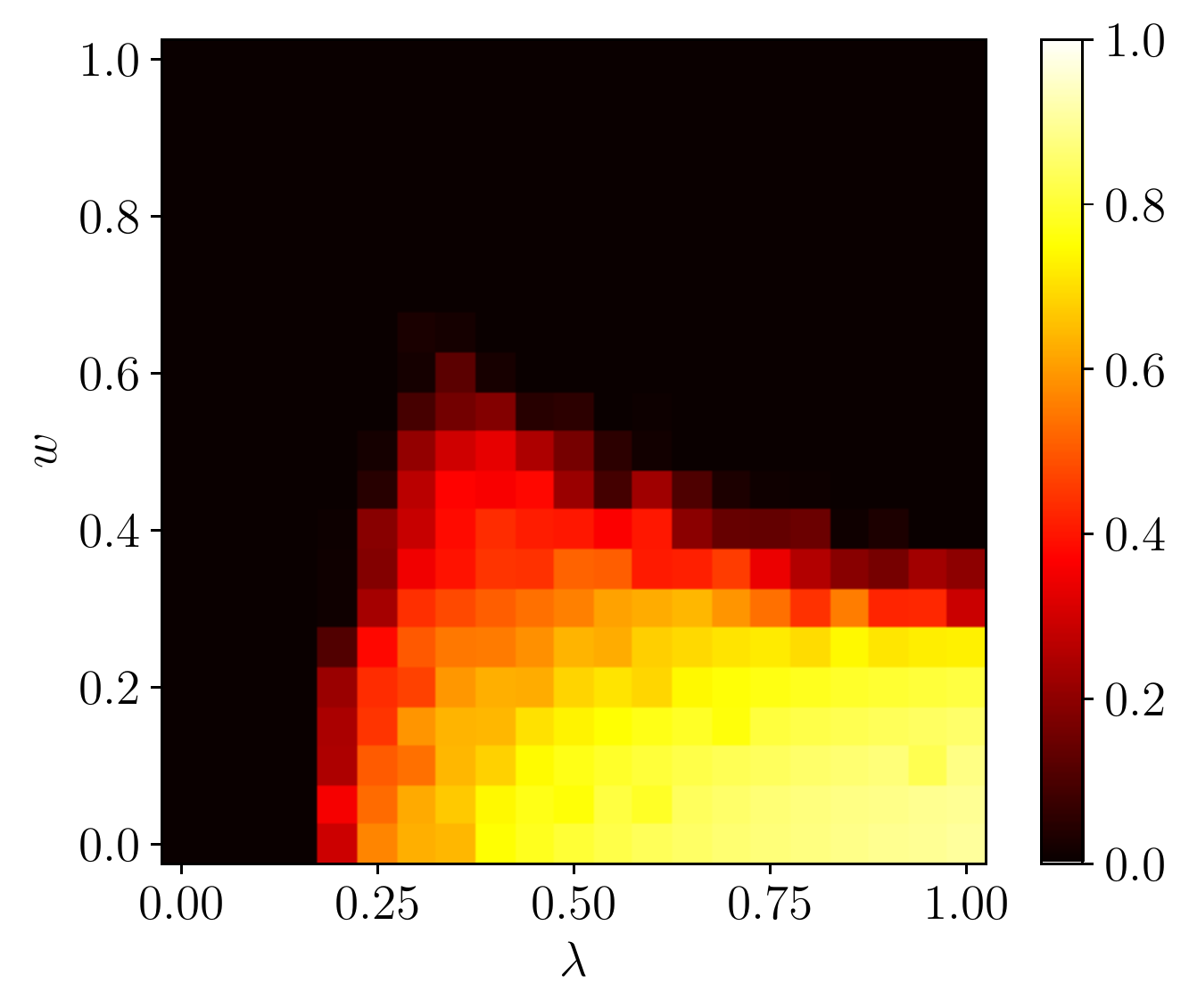}
    }
    \subfloat[$r=0.6$]{
        \includegraphics[width=0.32\textwidth]{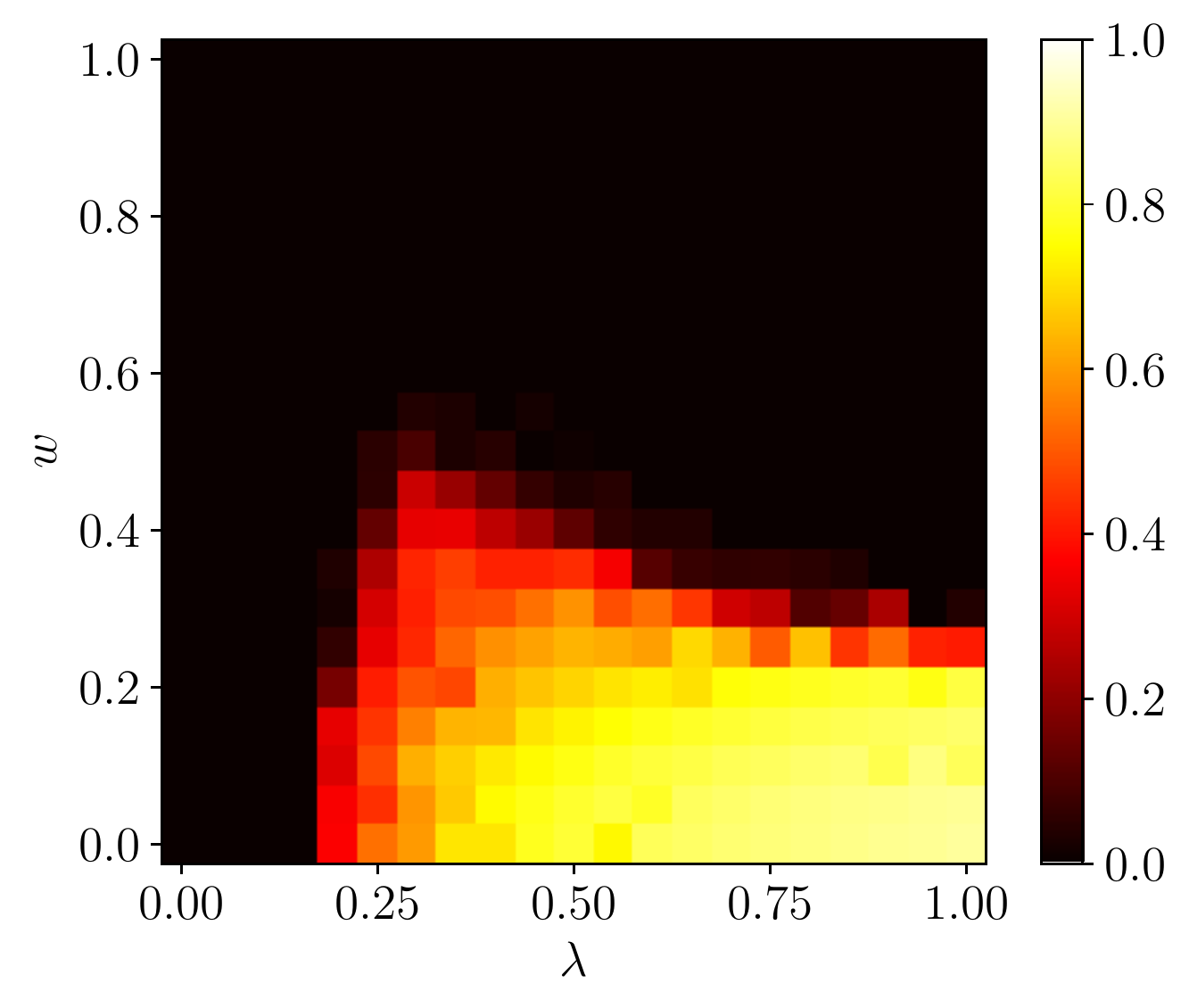}
    }
    \subfloat[$r=0.7$]{
        \includegraphics[width=0.32\textwidth]{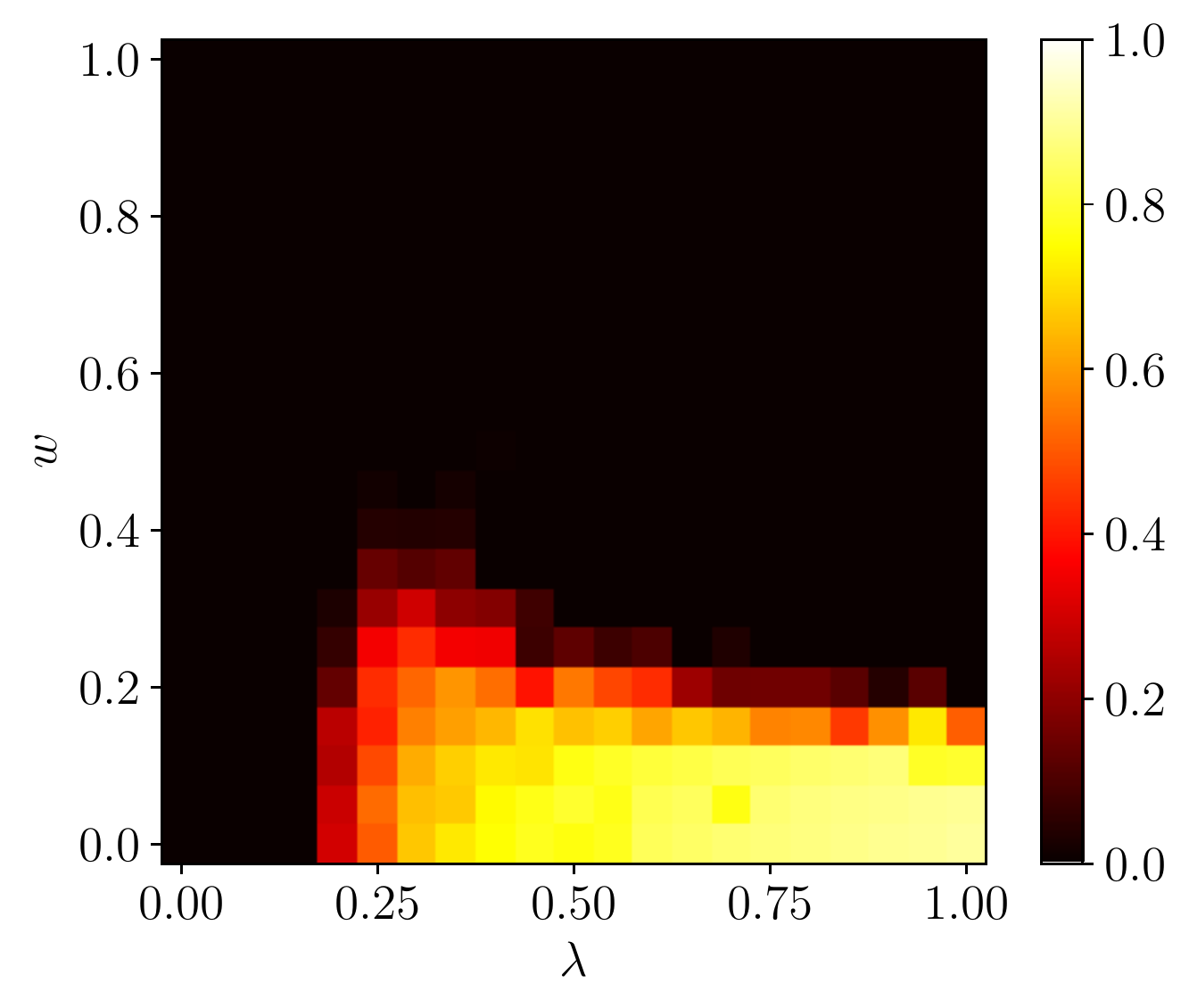}
    } \\
    \subfloat[$r=0.8$]{
        \includegraphics[width=0.32\textwidth]{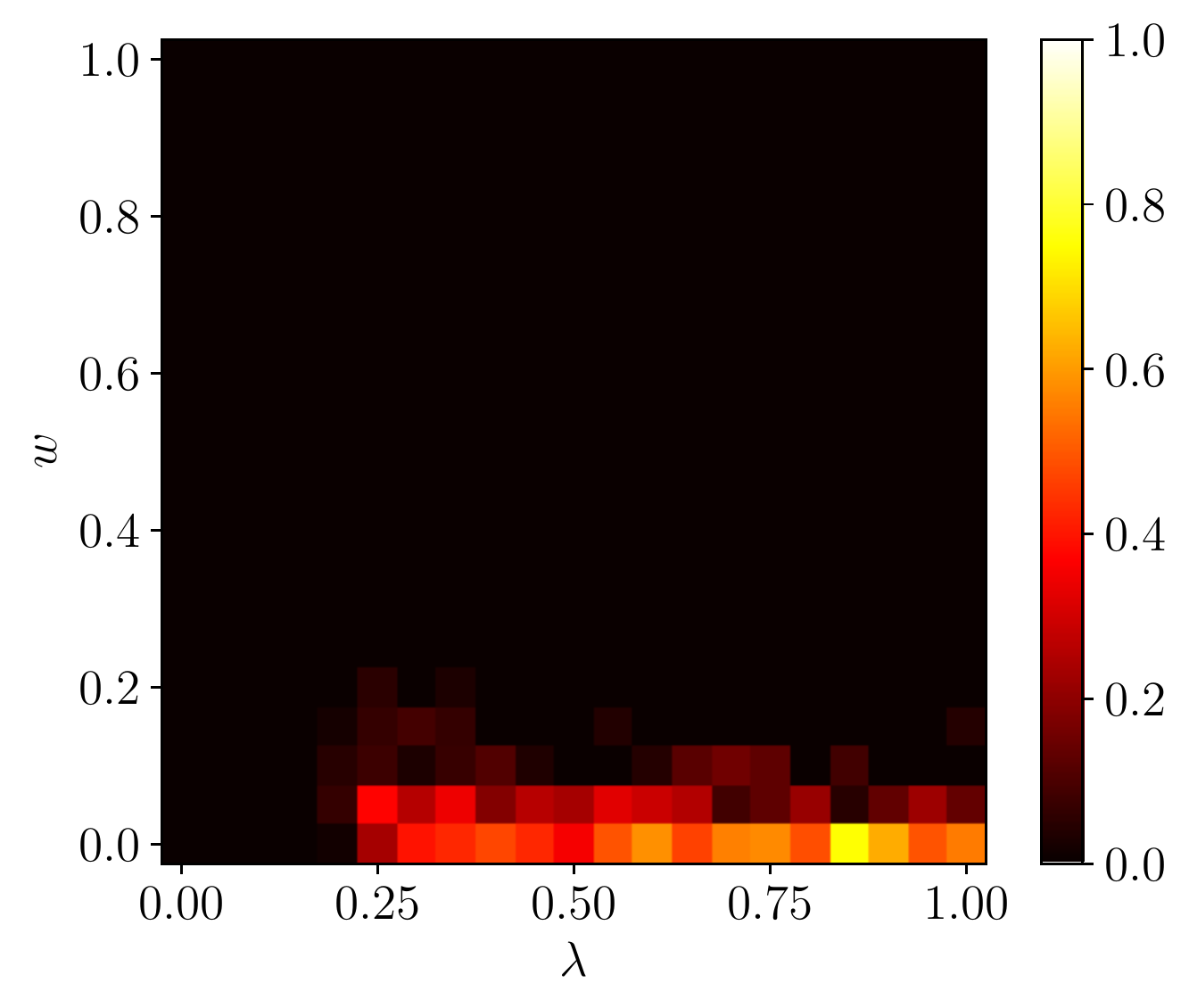}
    }
    \subfloat[$r=0.9$]{
        \includegraphics[width=0.32\textwidth]{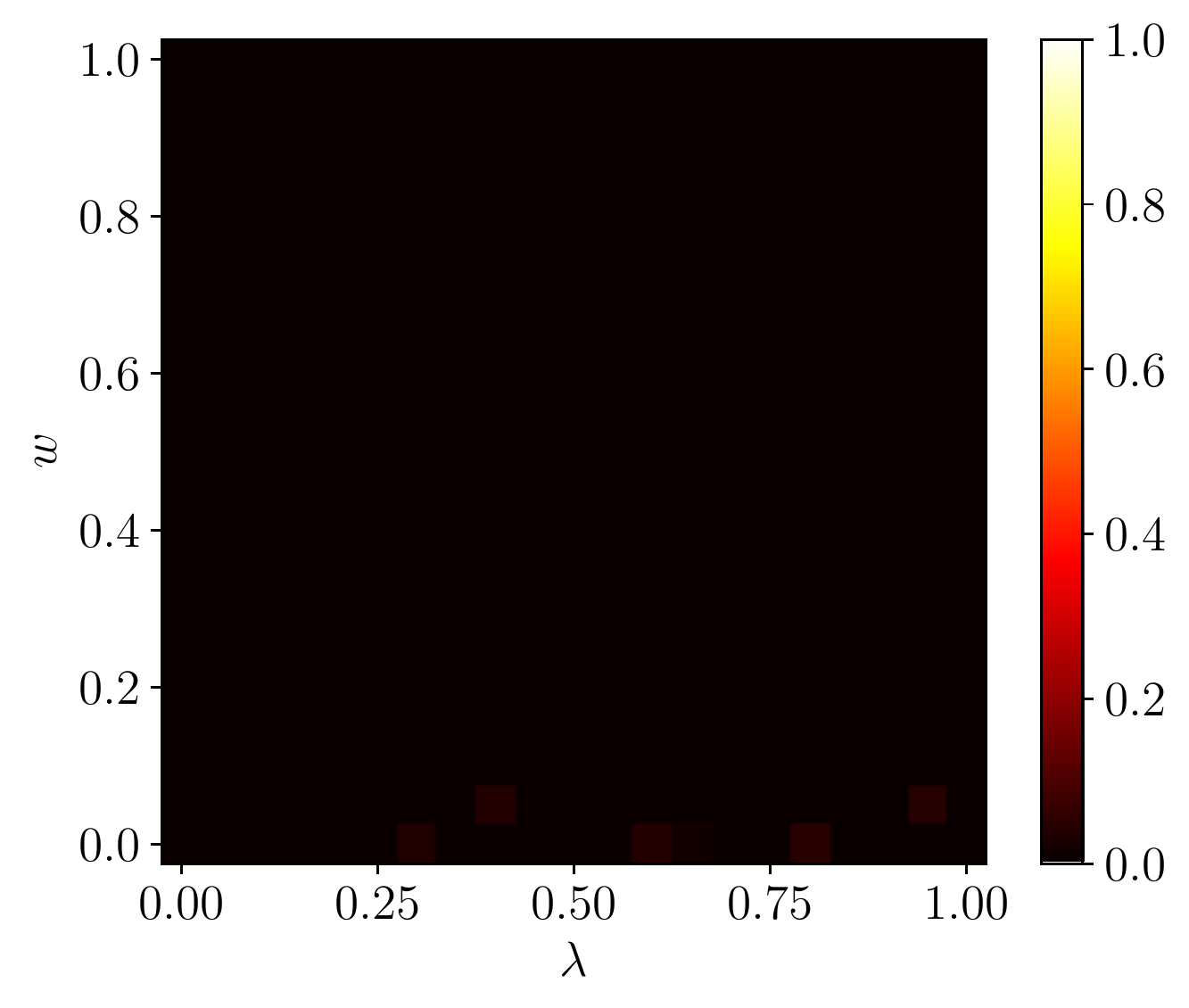}
    }
    \subfloat[$r=1.0$]{
        \includegraphics[width=0.32\textwidth]{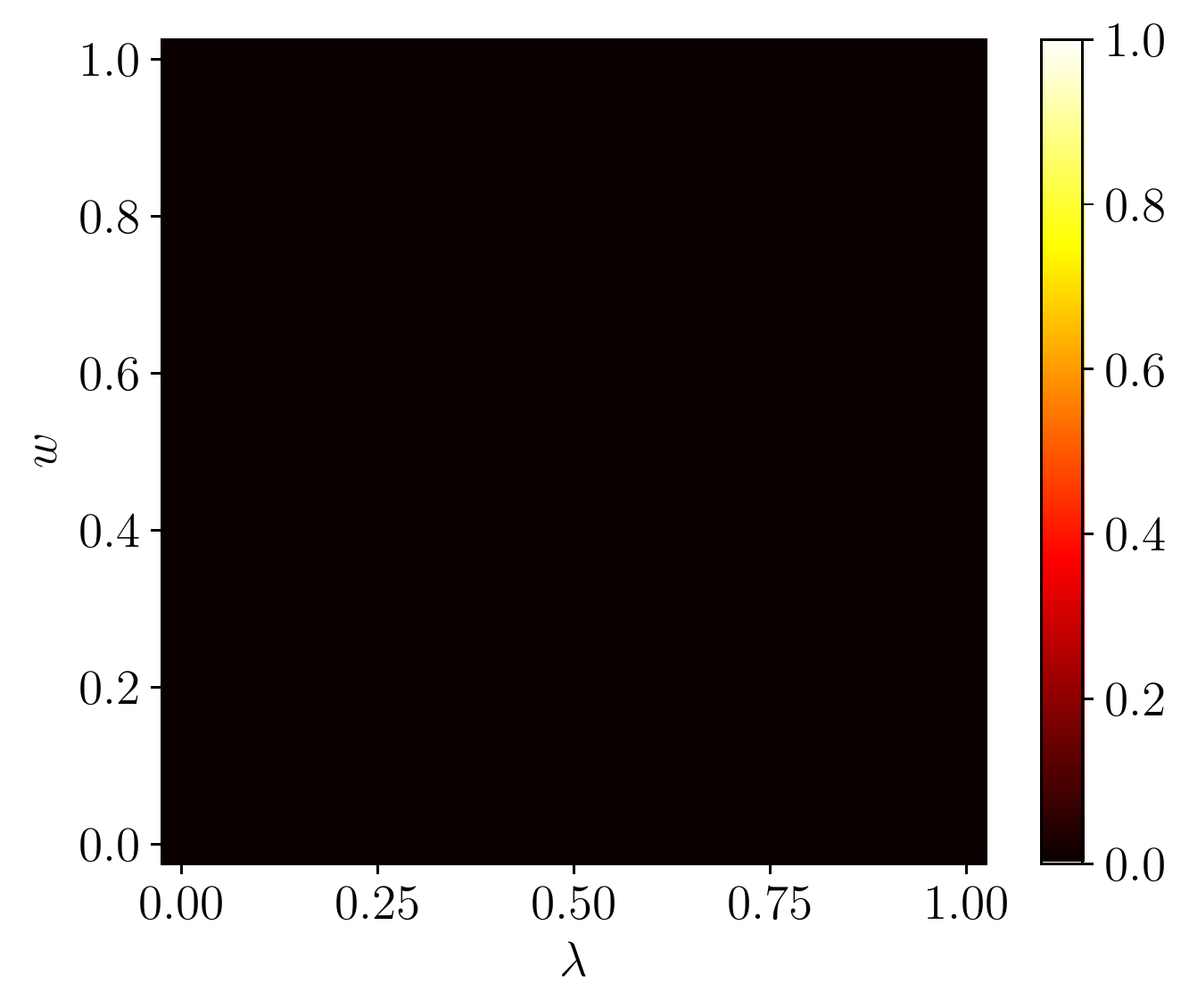}
    }
    \caption{Infected rate vs infection rate and risk perception for $D=0.2$ and various rewiring probabilities. }
    \label{IxWxL_D2}
\end{figure}

In the long term this effect seems to be mostly positive, as can be seen in \cref{IxWxL_D2} were we see the average infection rate in the stationary state. In the long term mild vaccine coverage seems to be sufficient to stop the epidemic spreading as the endemic region is reduced as $r$ increases. This effect is even more pronounced when there is more rewiring as we will see in the next section.

\FloatBarrier
\subsection{$D=0.4$}

For $D=0.4$, the agents start with a stronger degree of disagreement. This facilitates the rewiring process making rewiring more likely. We will now observe that the effects mentioned in the previous section are reinforced.

\begin{figure}[h]
    \centering
    \subfloat[$r=0.5$]{
        \includegraphics[width=0.32\textwidth]{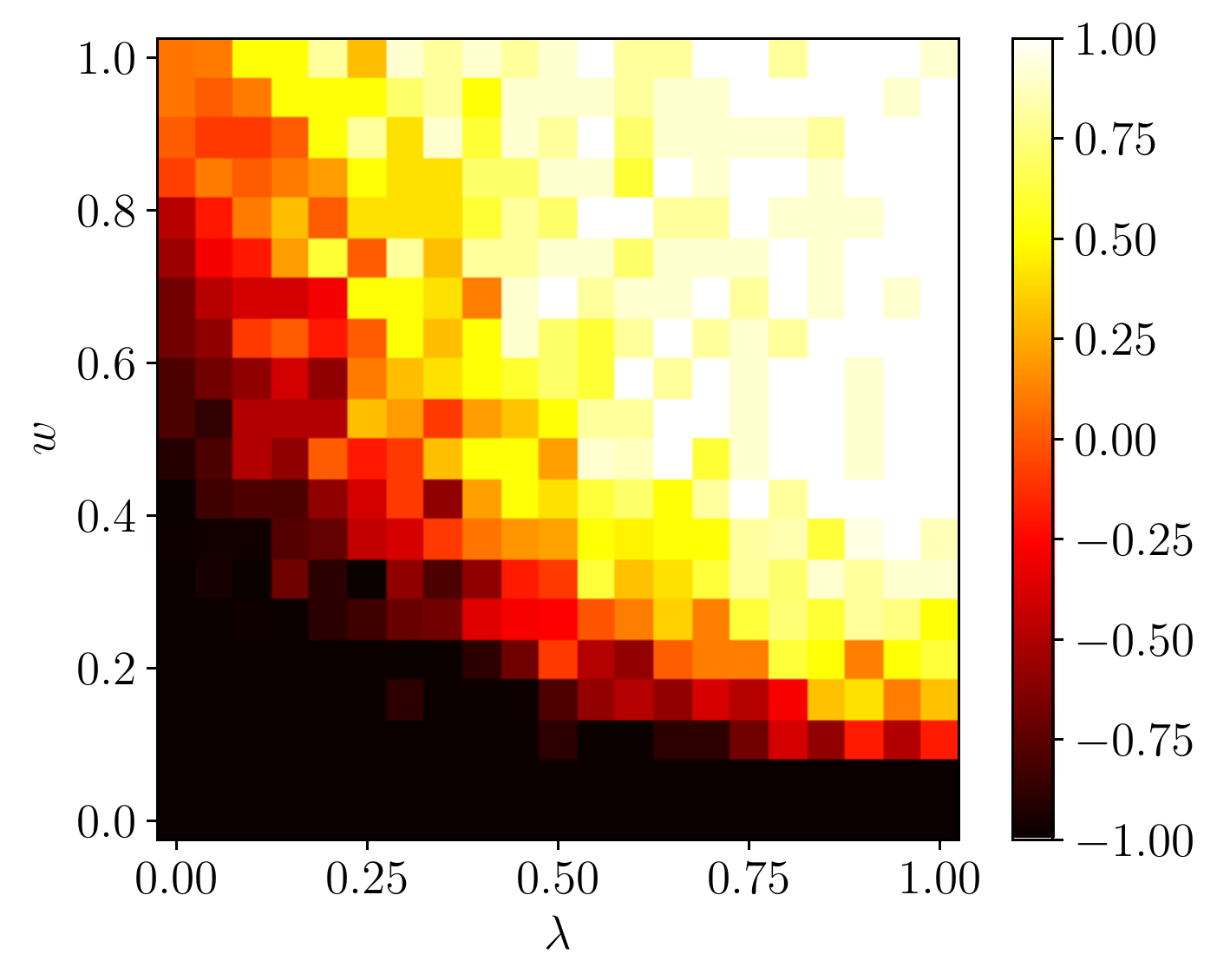}
    }
    \subfloat[$r=0.6$]{
        \includegraphics[width=0.32\textwidth]{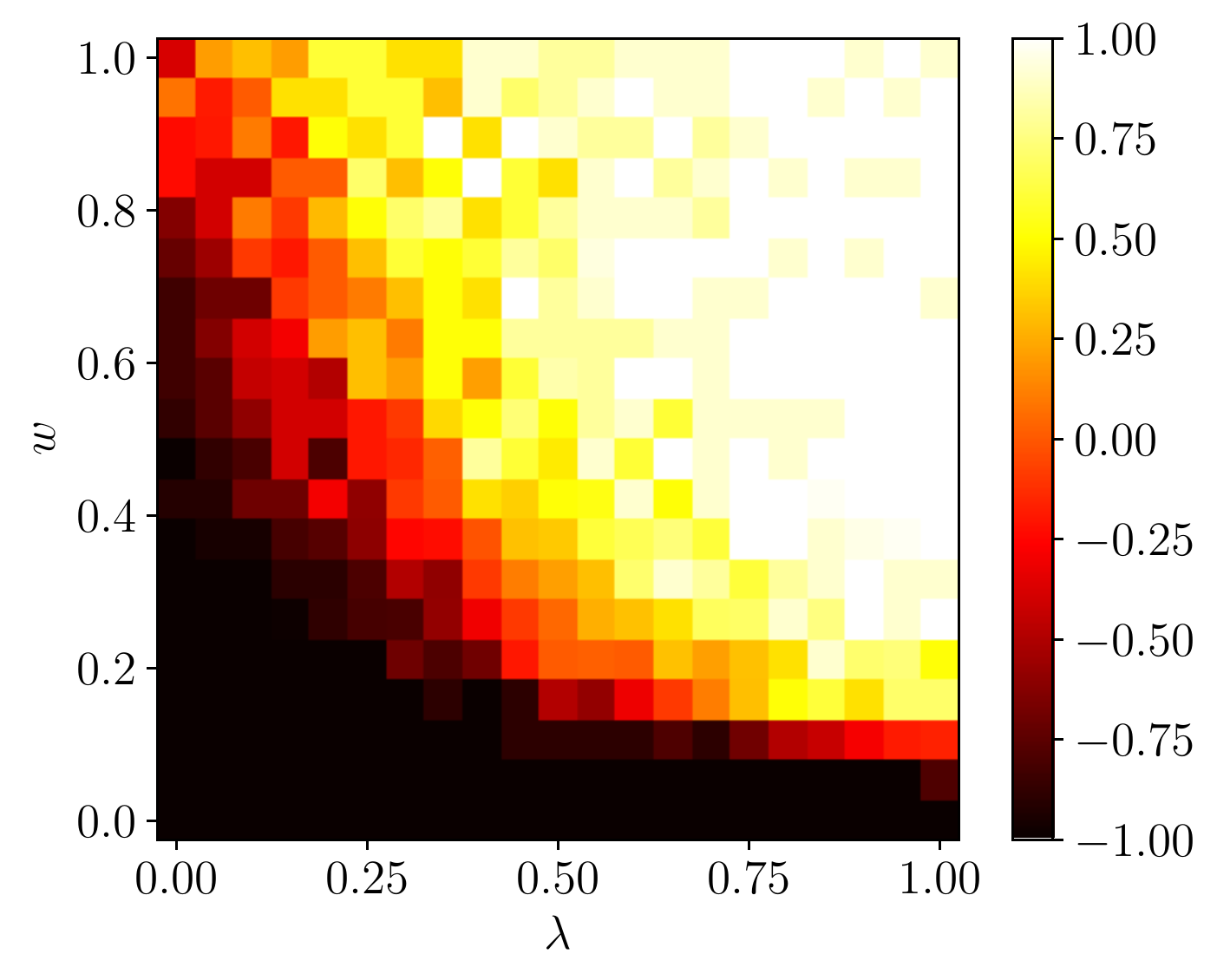}
    }
    \subfloat[$r=0.7$]{
        \includegraphics[width=0.32\textwidth]{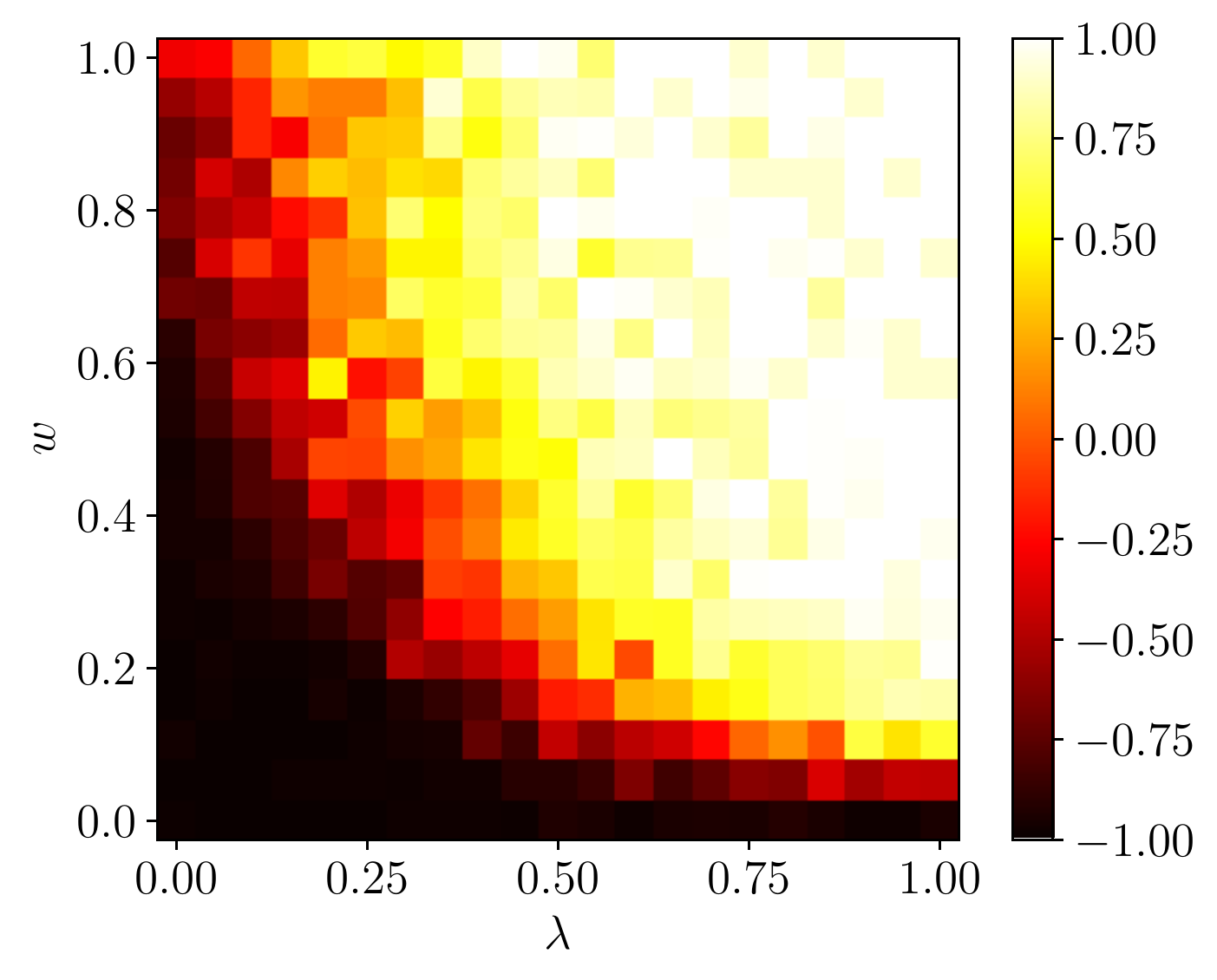}
    } \\
    \subfloat[$r=0.8$]{
        \includegraphics[width=0.32\textwidth]{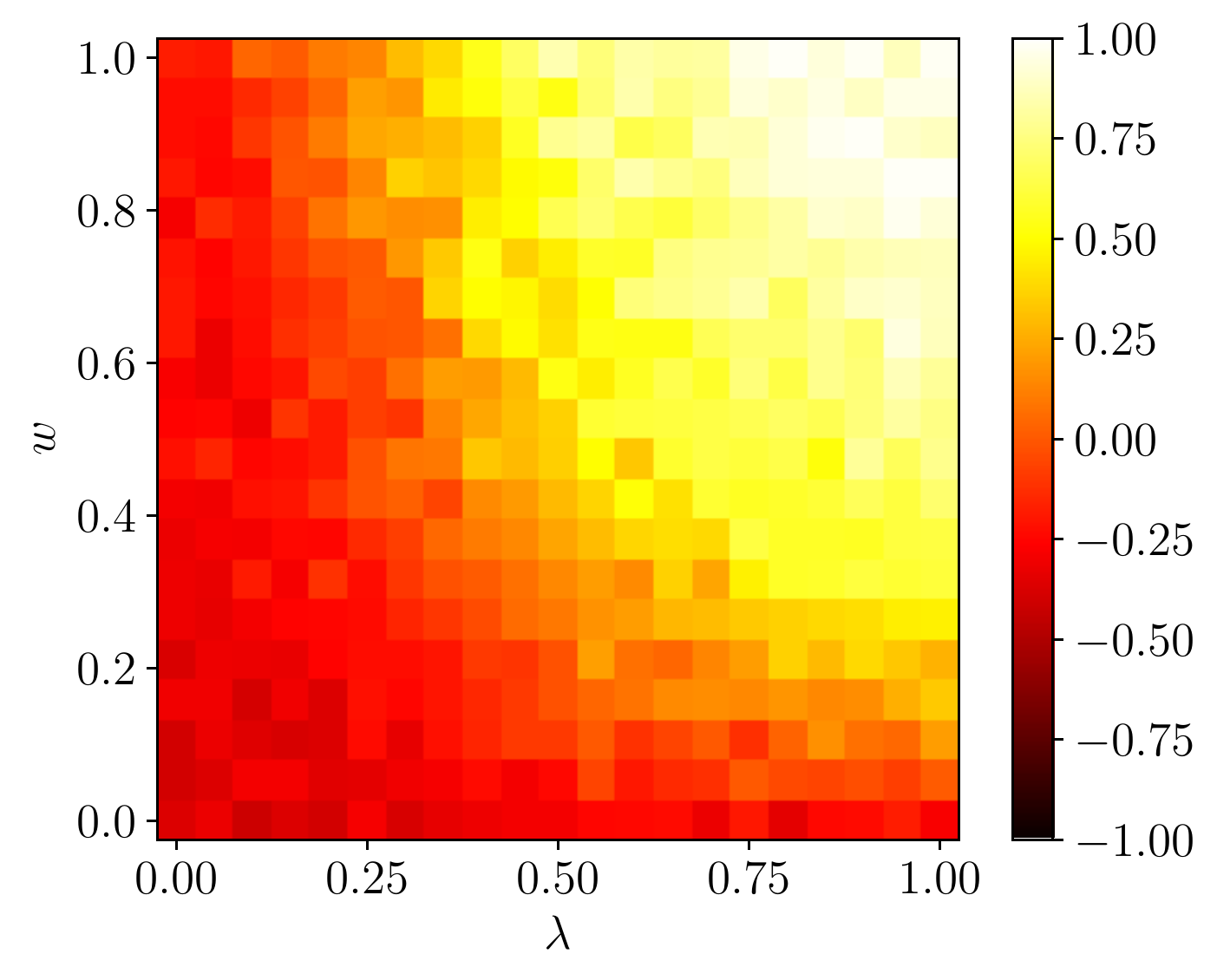}
    }
    \subfloat[$r=0.9$]{
        \includegraphics[width=0.32\textwidth]{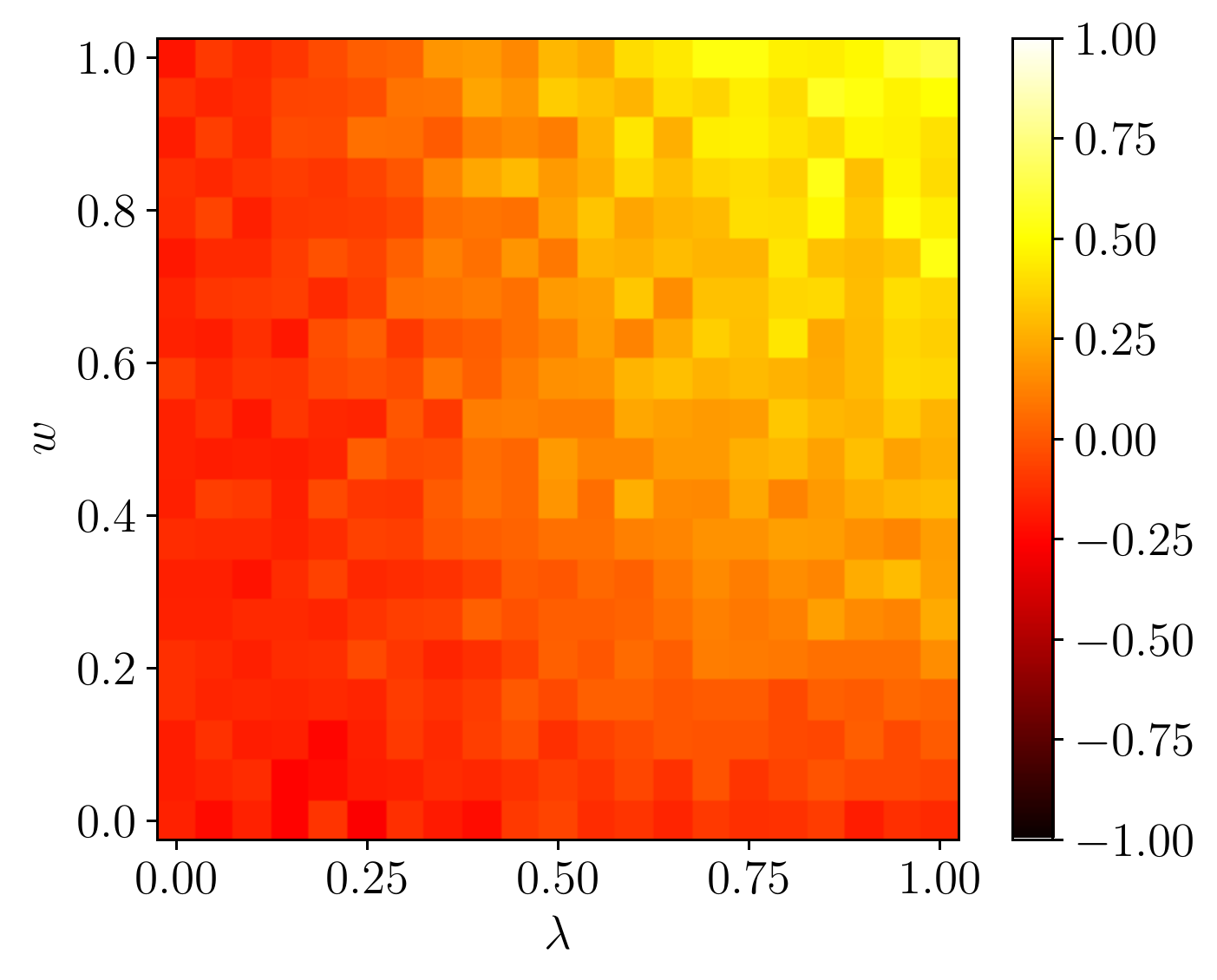}
    }
    \subfloat[$r=1.0$]{
        \includegraphics[width=0.32\textwidth]{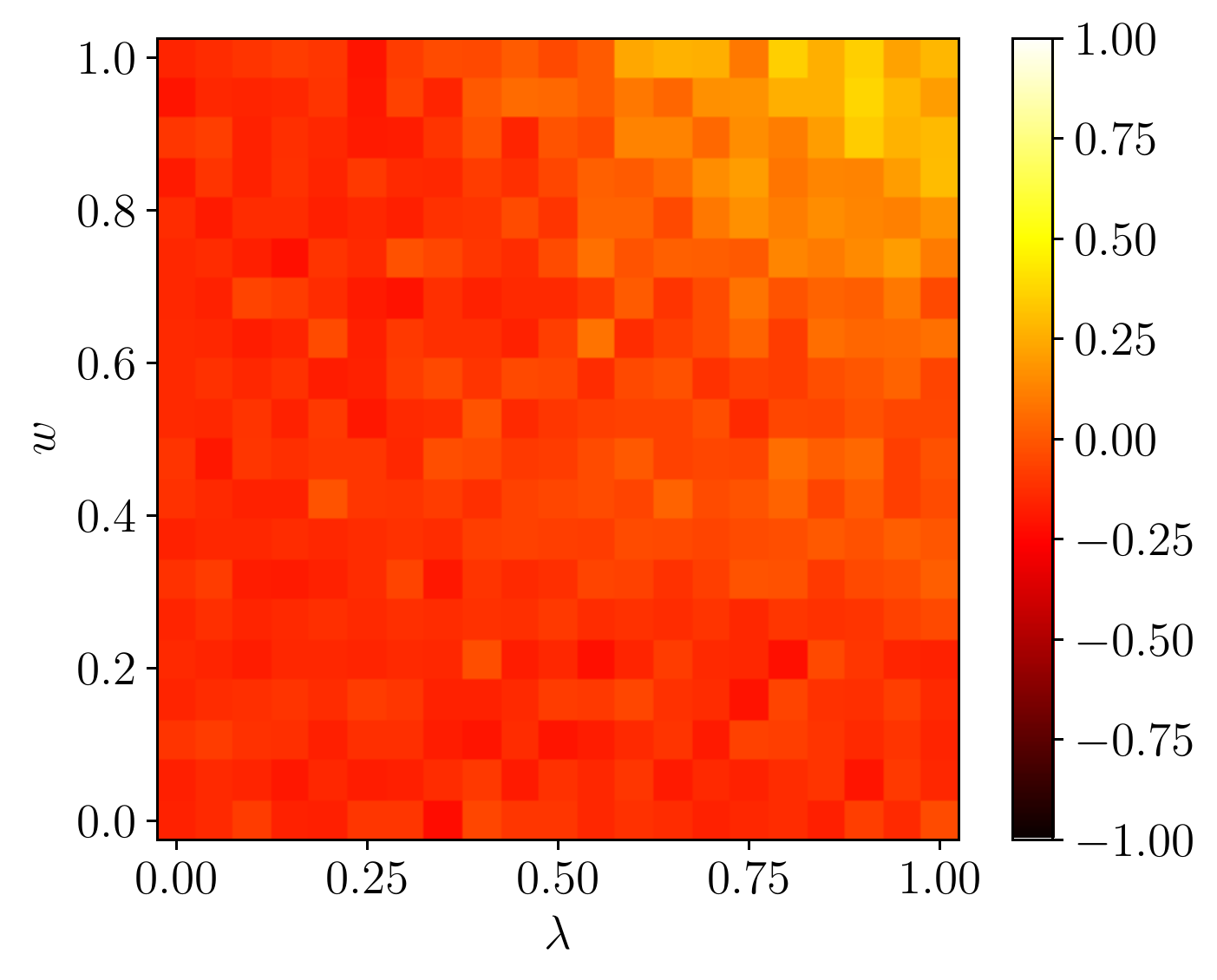}
    }
    \caption{Average opinion for different parameter configurations}
    \label{OxWxL_D4}
\end{figure}

\begin{figure}[h]
    \centering
    \subfloat[$r=0.5$]{
        \includegraphics[width=0.32\textwidth]{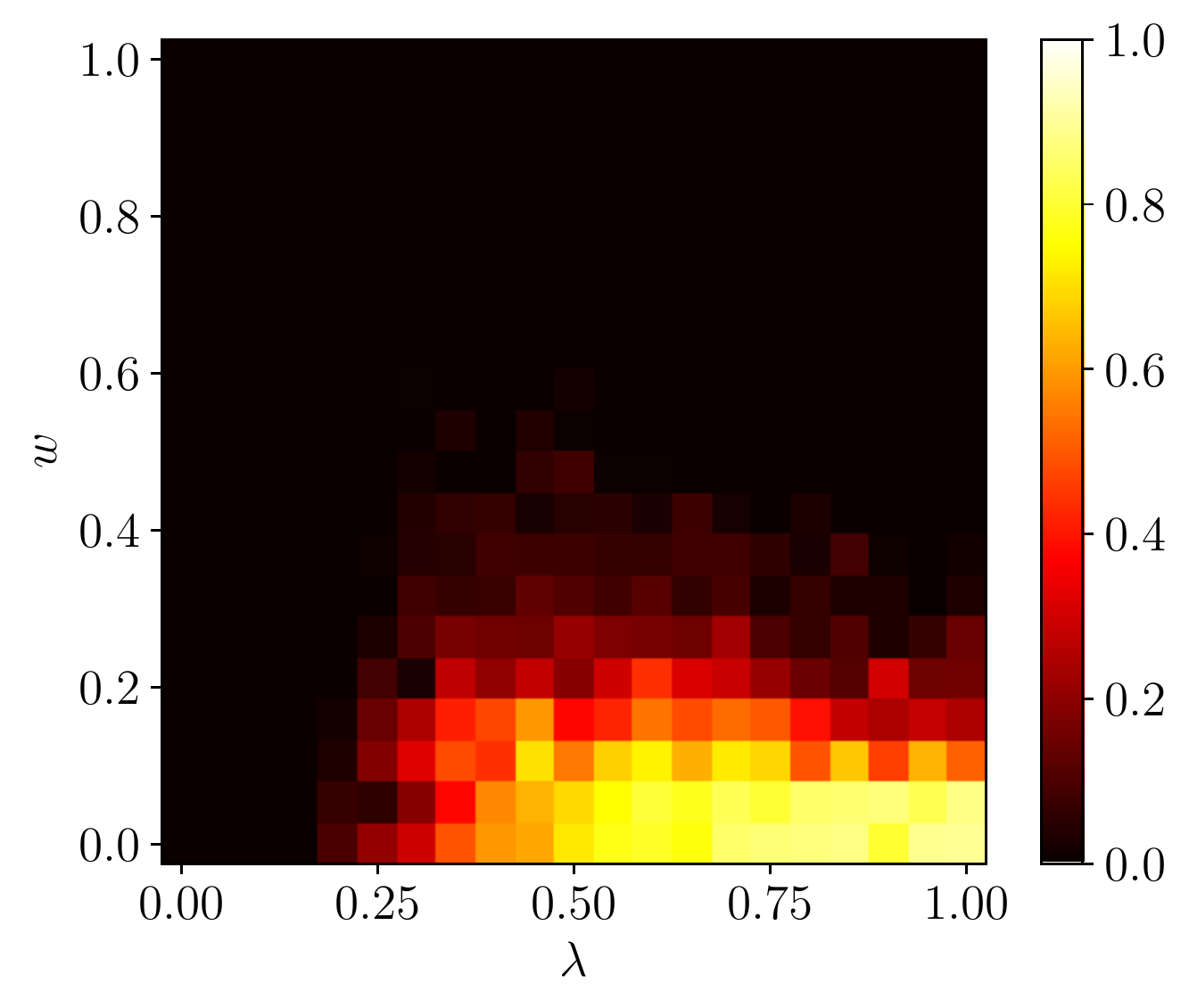}
    }
    \subfloat[$r=0.6$]{
        \includegraphics[width=0.32\textwidth]{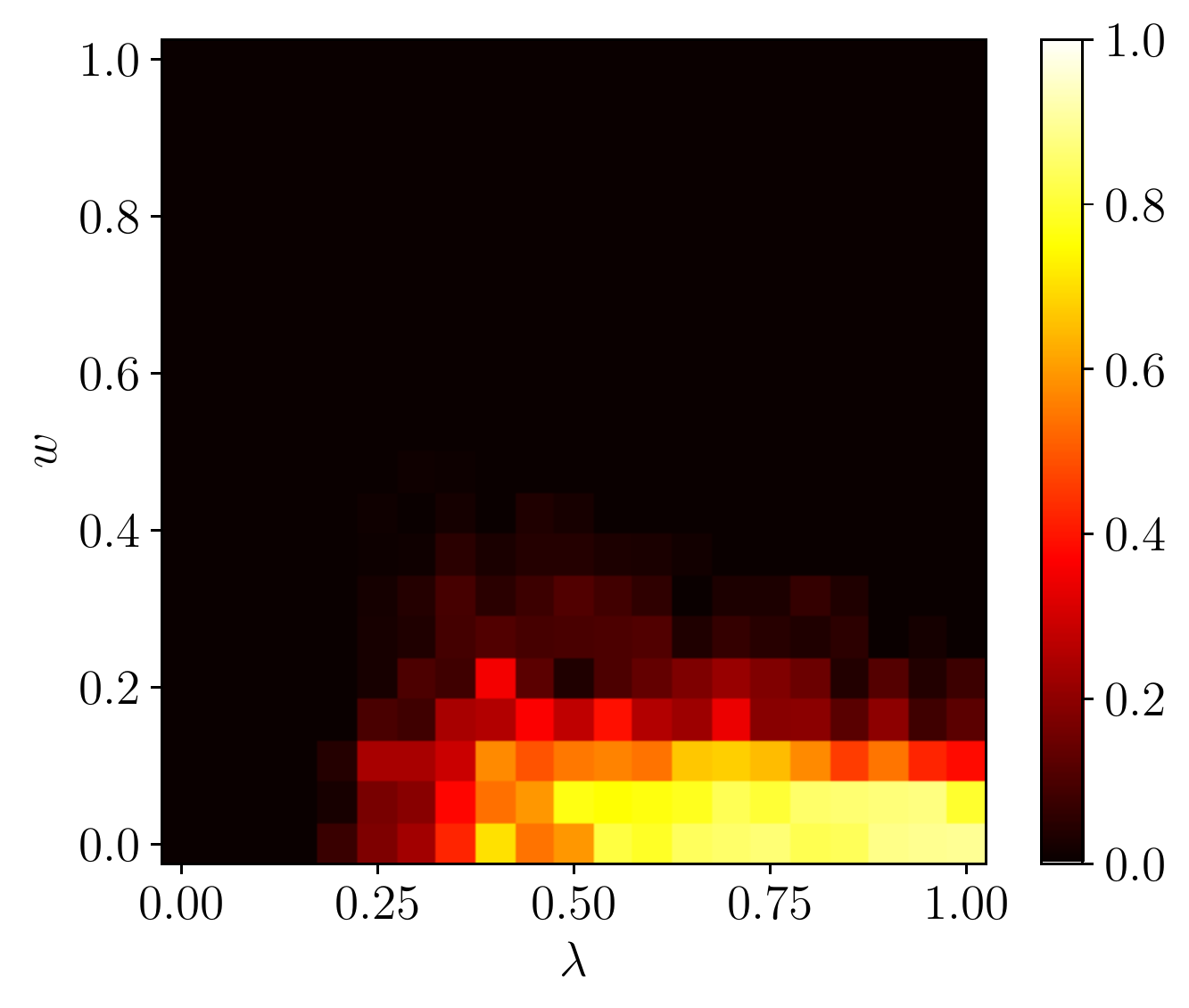}
    }
    \subfloat[$r=0.7$]{
        \includegraphics[width=0.32\textwidth]{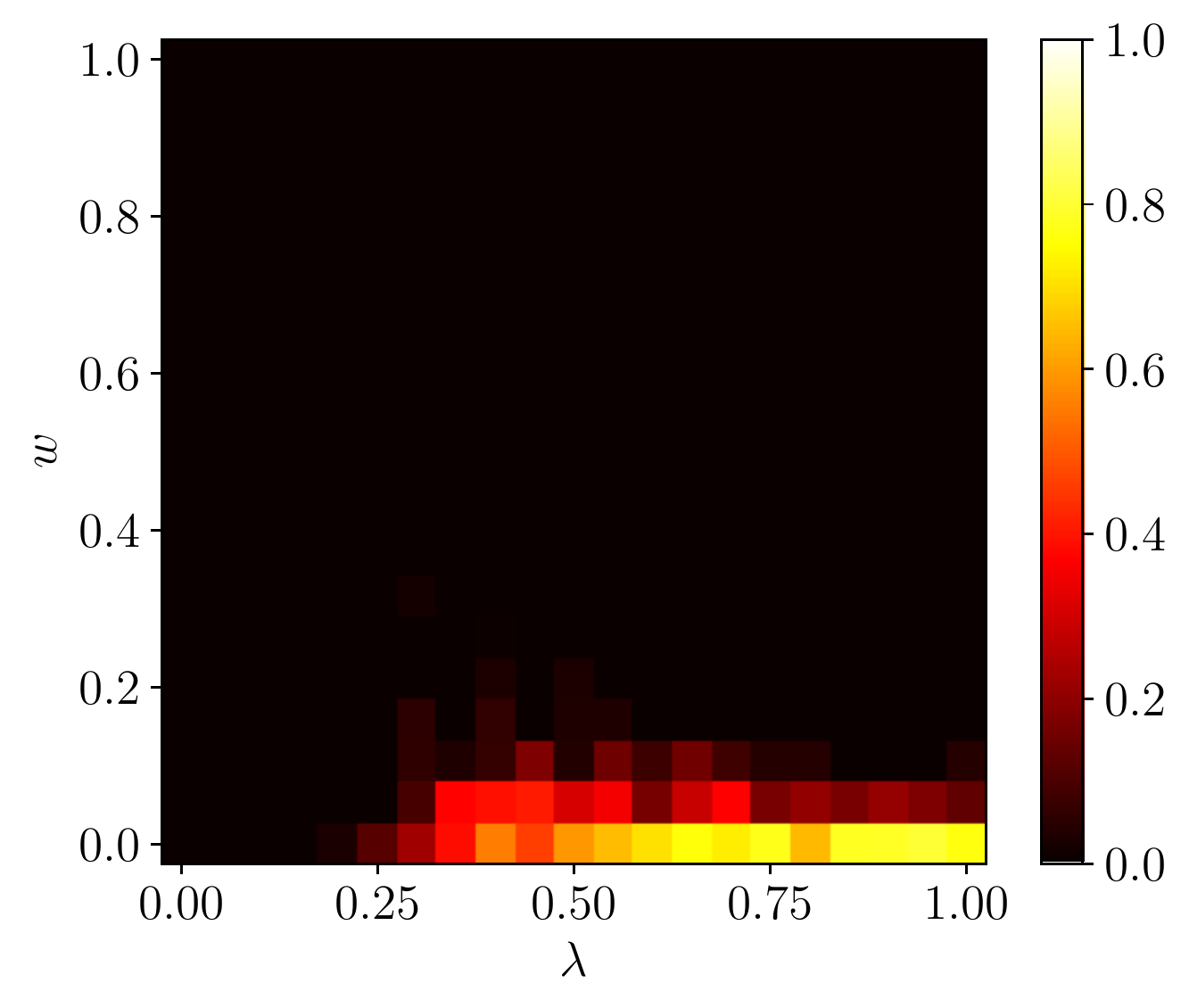}
    } \\
    \subfloat[$r=0.8$]{
        \includegraphics[width=0.32\textwidth]{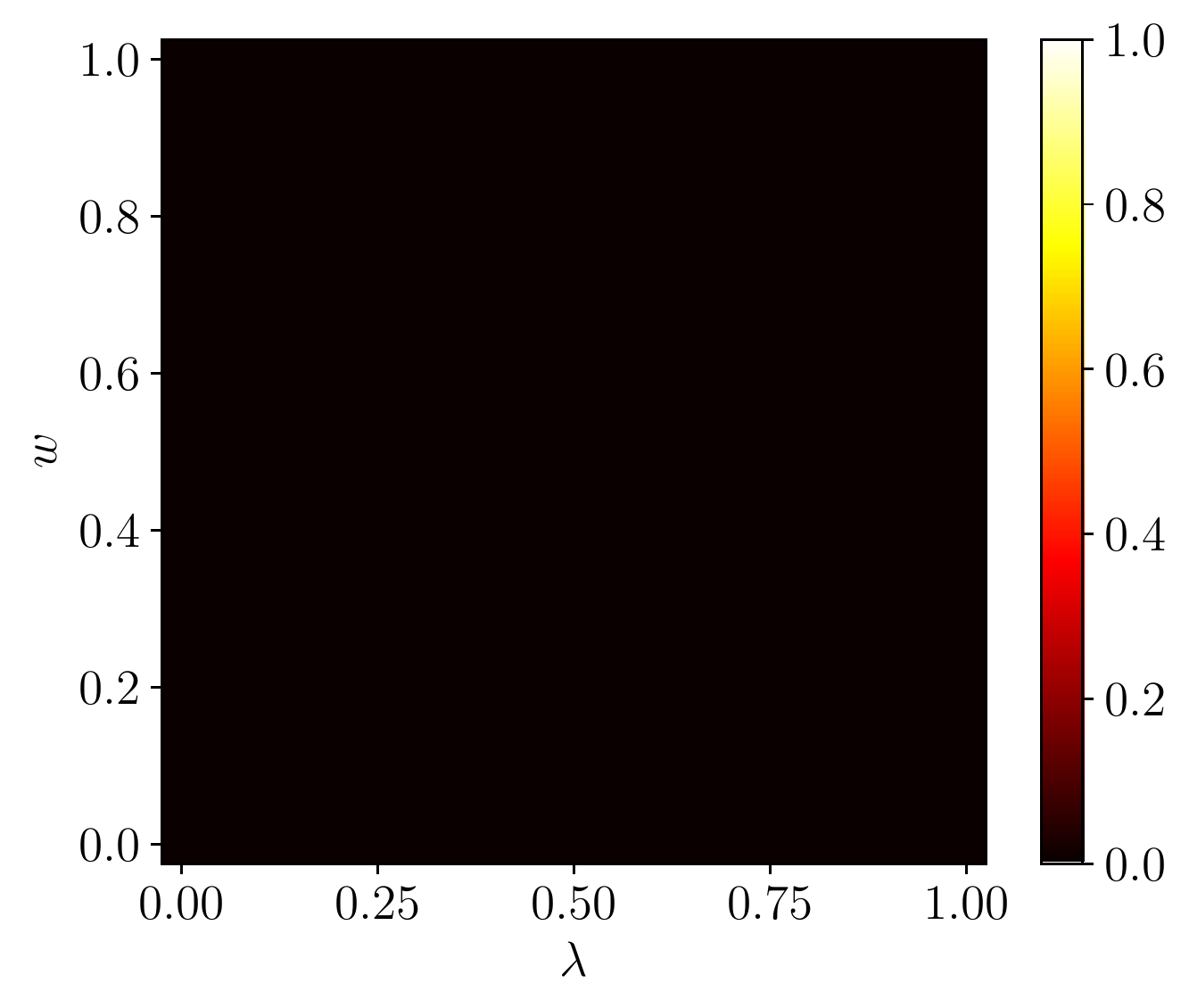}
    }
    \subfloat[$r=0.9$]{
        \includegraphics[width=0.32\textwidth]{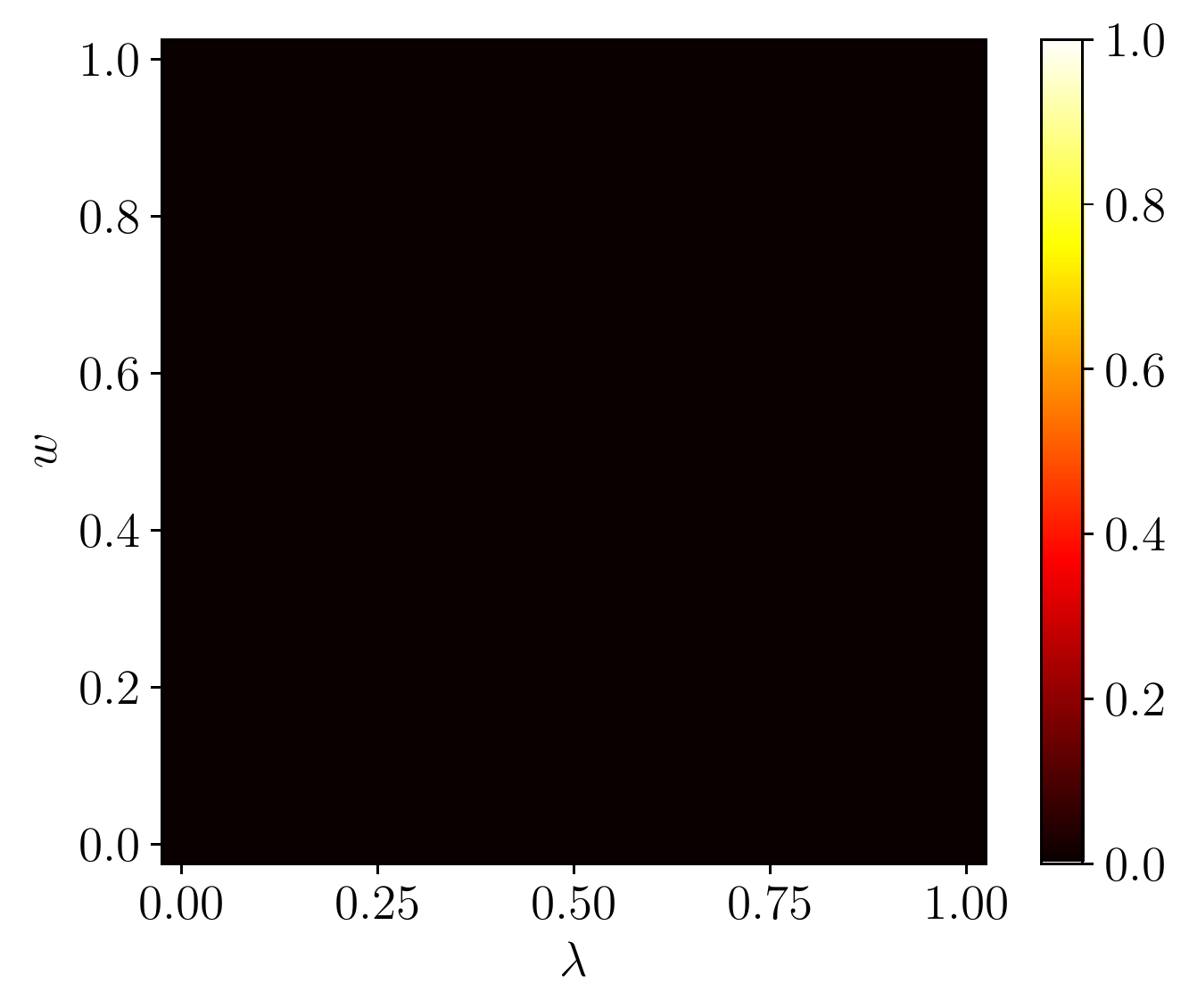}
    }
    \subfloat[$r=1.0$]{
        \includegraphics[width=0.32\textwidth]{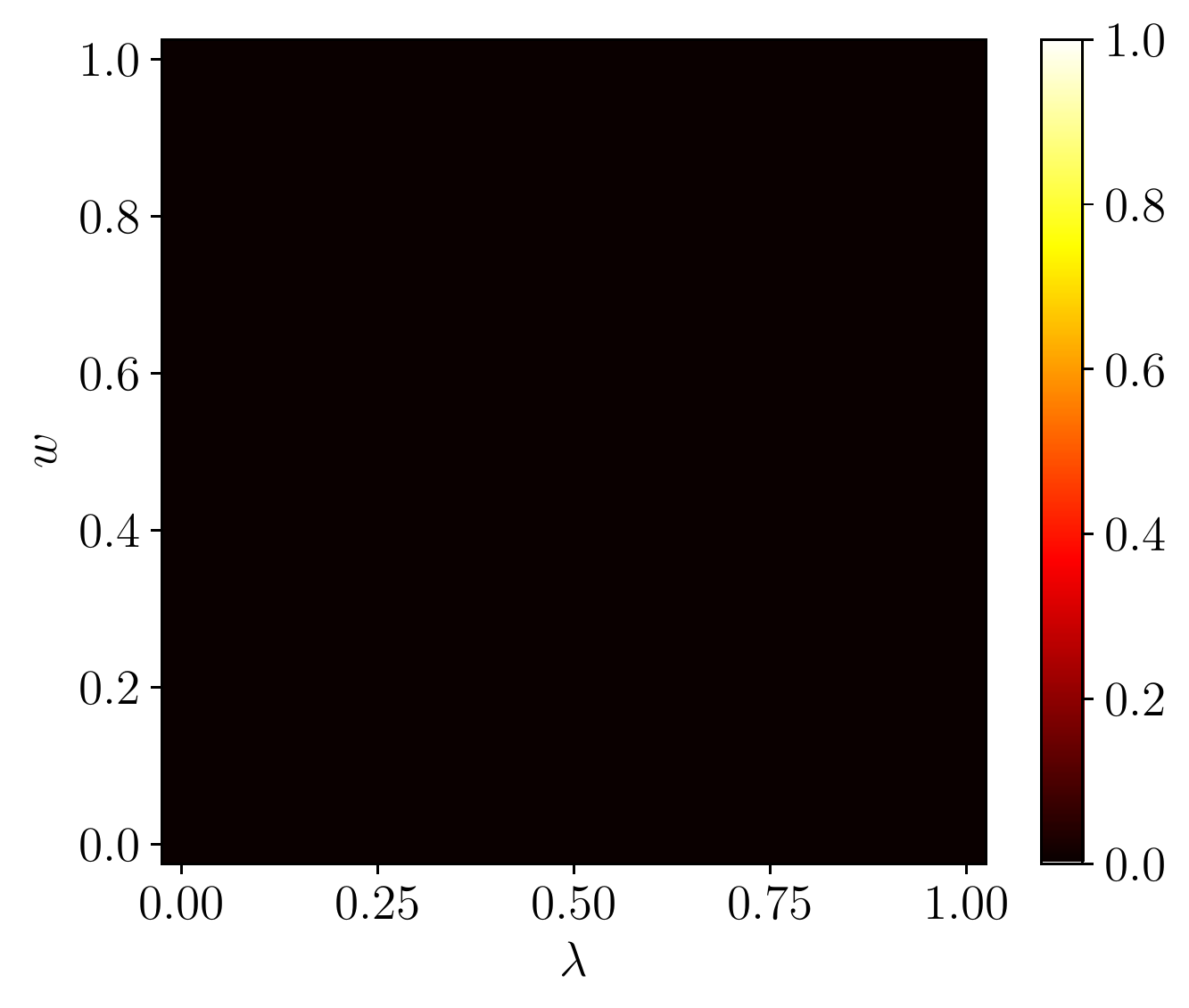}
    }
    \caption{Infected rate for different parameter configurations}
    \label{IxWxL_D4}
\end{figure}

In \cref{OxWxL_D4}, we see that for $r \approx 0.6$, the region with a pro-vaccine opinion increases. It is striking that as the probability of rewiring increases, the anti-vaccination opinion is preserved even though the epidemic spreading vanishes, as shown in \cref{IxWxL_D4}. On the one hand, rewiring preserves pockets of anti-vaccination individuals, while on other hand, hinders the epidemic spreading. This suggests that with a higher probability of rewiring there is a smaller vaccine coverage.

\begin{figure}[h]
    \centering
    \subfloat[Steady state infection rate.]{
        \includegraphics[width=0.32\textwidth]{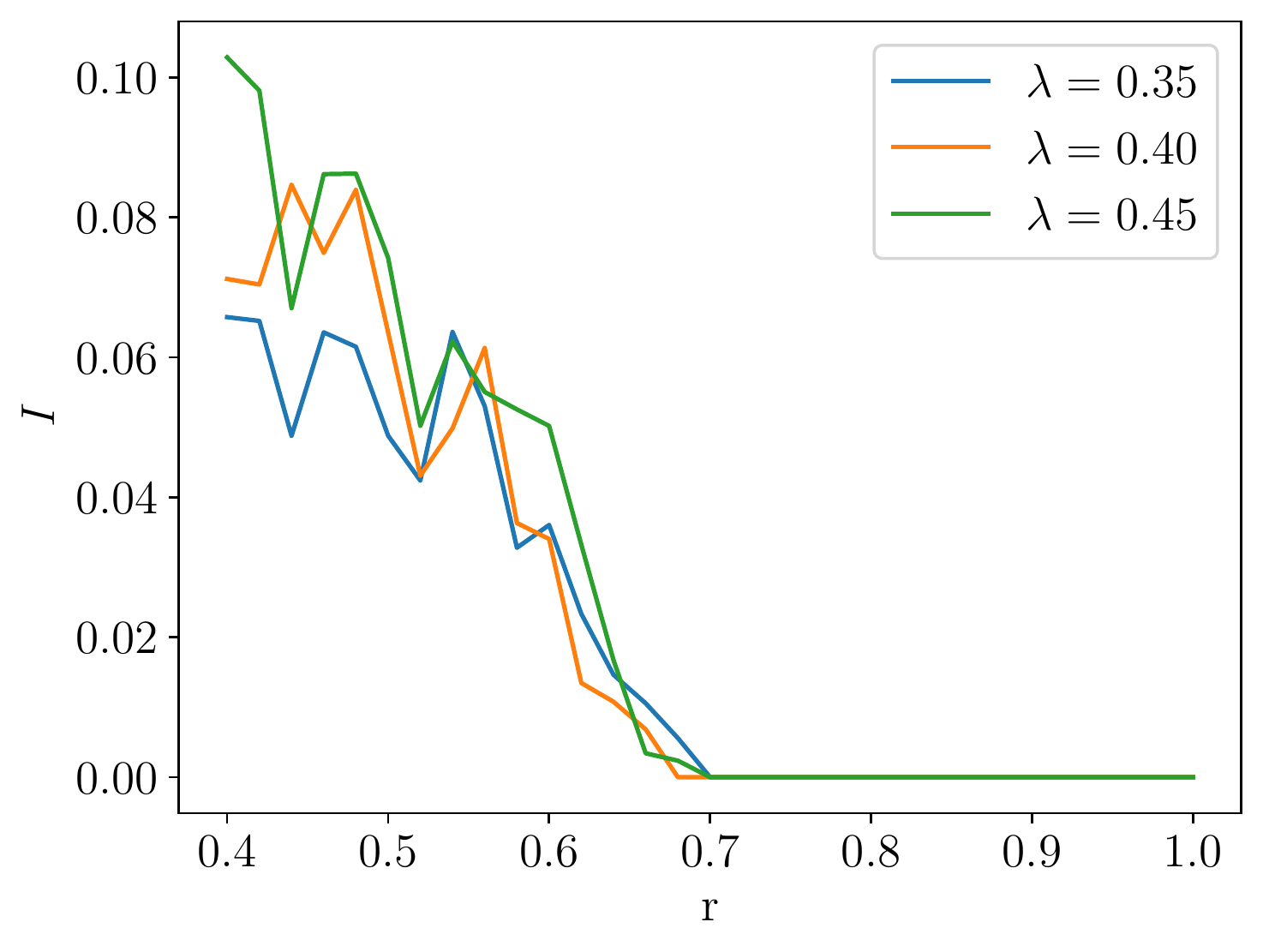}
    }
    \subfloat[Peak infection rate.]{
        \includegraphics[width=0.32\textwidth]{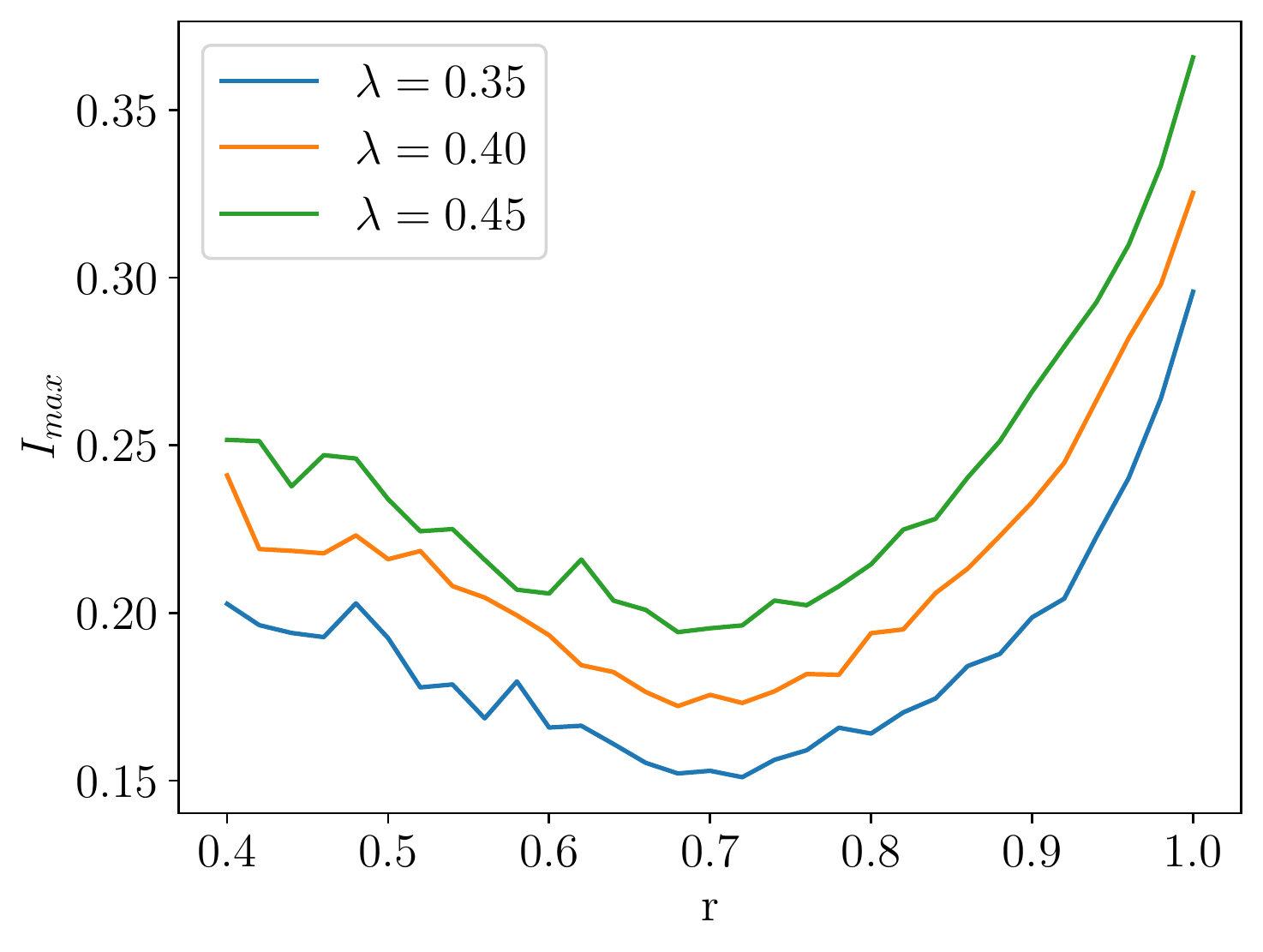}
    }
    \subfloat[Average number of disconnected networks.]{
        \includegraphics[width=0.32\textwidth]{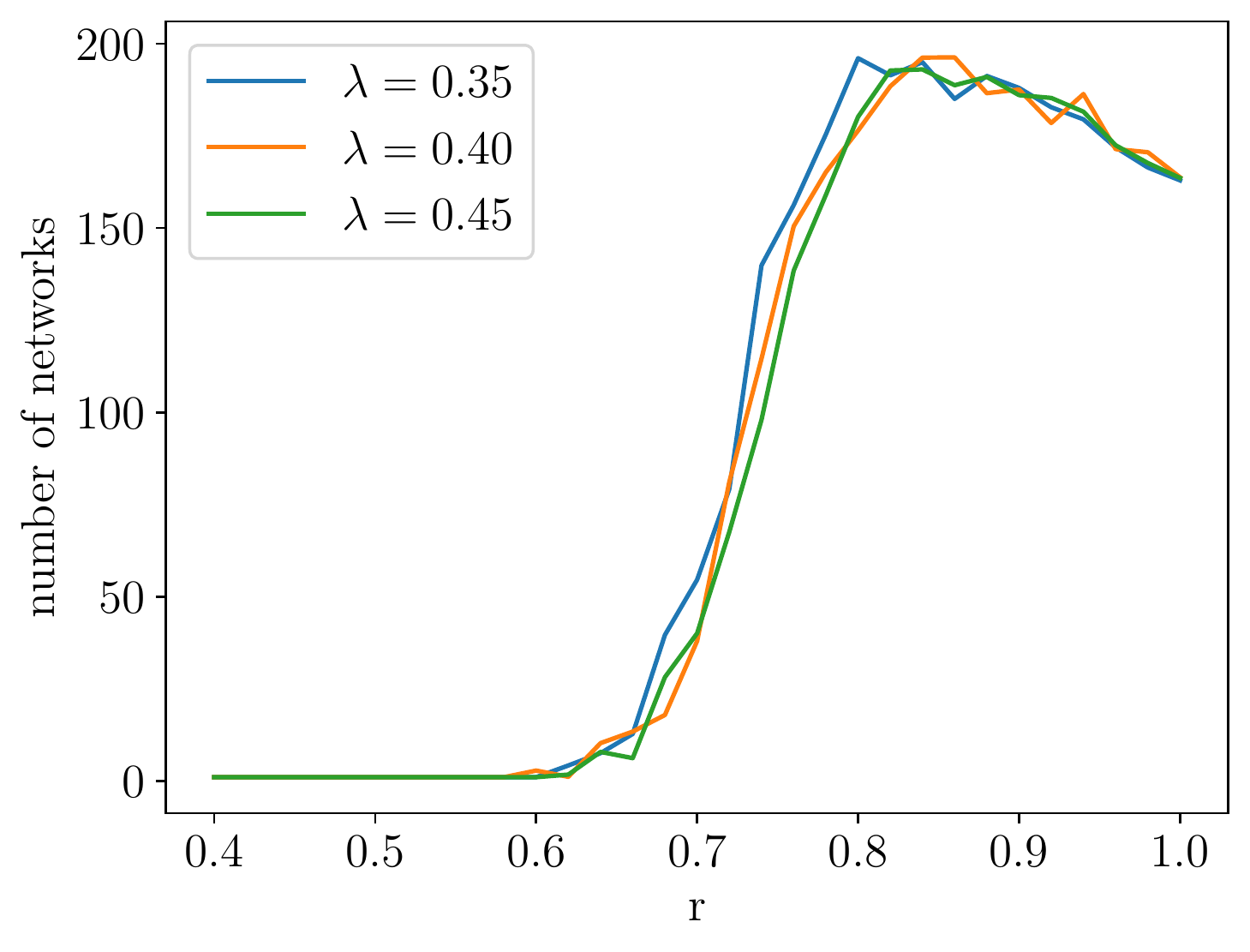}
    }
    \caption{Value of the steady state infection rate in (a), maximum infection rate over time in (b) and number of disconnected networks in the steady state in (c) versus rewiring for different values of $\lambda$, $w=0.4$ and $D=0.4$. Here we can see that the value of the epidemic peak reaches a minimum for $r \approx 0.7$ and then starts increasing.}
    \label{ImaxR}
\end{figure}

With a lower vaccine coverage, higher values of rewiring tend not to keep the epidemic spreading in the short time under check. This can be observed in \cref{ImaxR}, where we examine the value of the infection peak, i.e., the maximum number of simultaneously infected agents. In this figure, we can observe that this value drops until $r \approx 0.7$, where it starts to increase again.

For public policy, one of the most relevant factors to consider is the number of simultaneously infected agents. This is because the healthcare system can only provide care to a certain number of patients, which is usually a small fraction of the population. For this reason, it is crucial to keep the infected fraction of the population bellow the relative capacity of the healthcare system.

The optimal rewiring occurs when the network starts to break apart. When anti-vaccine agents become segregated, they become especially susceptible to the spread of the disease. A high degree of non-vaccinated individuals connected results in faster disease spread and, therefore, an increase in the value of the epidemic peak. To prevent the epidemic peak from increasing, a complete separation between pro and anti-vaccine groups is undesirable.

\begin{figure}[hb]
    \centering
    \subfloat[$r=0.50$]{
        \includegraphics[width=0.32\textwidth]{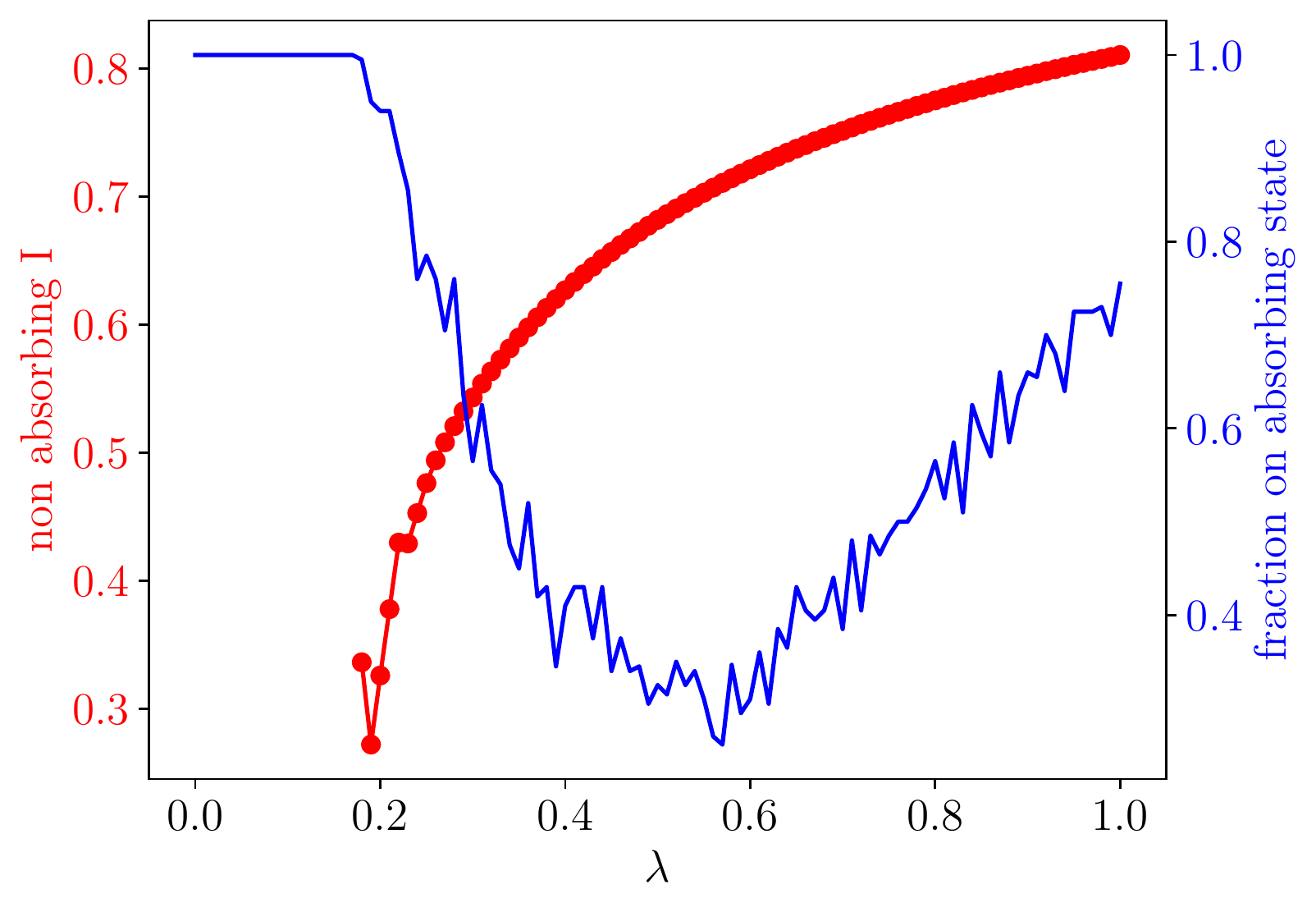}
    }
    \subfloat[]{
        \includegraphics[width=0.32\textwidth]{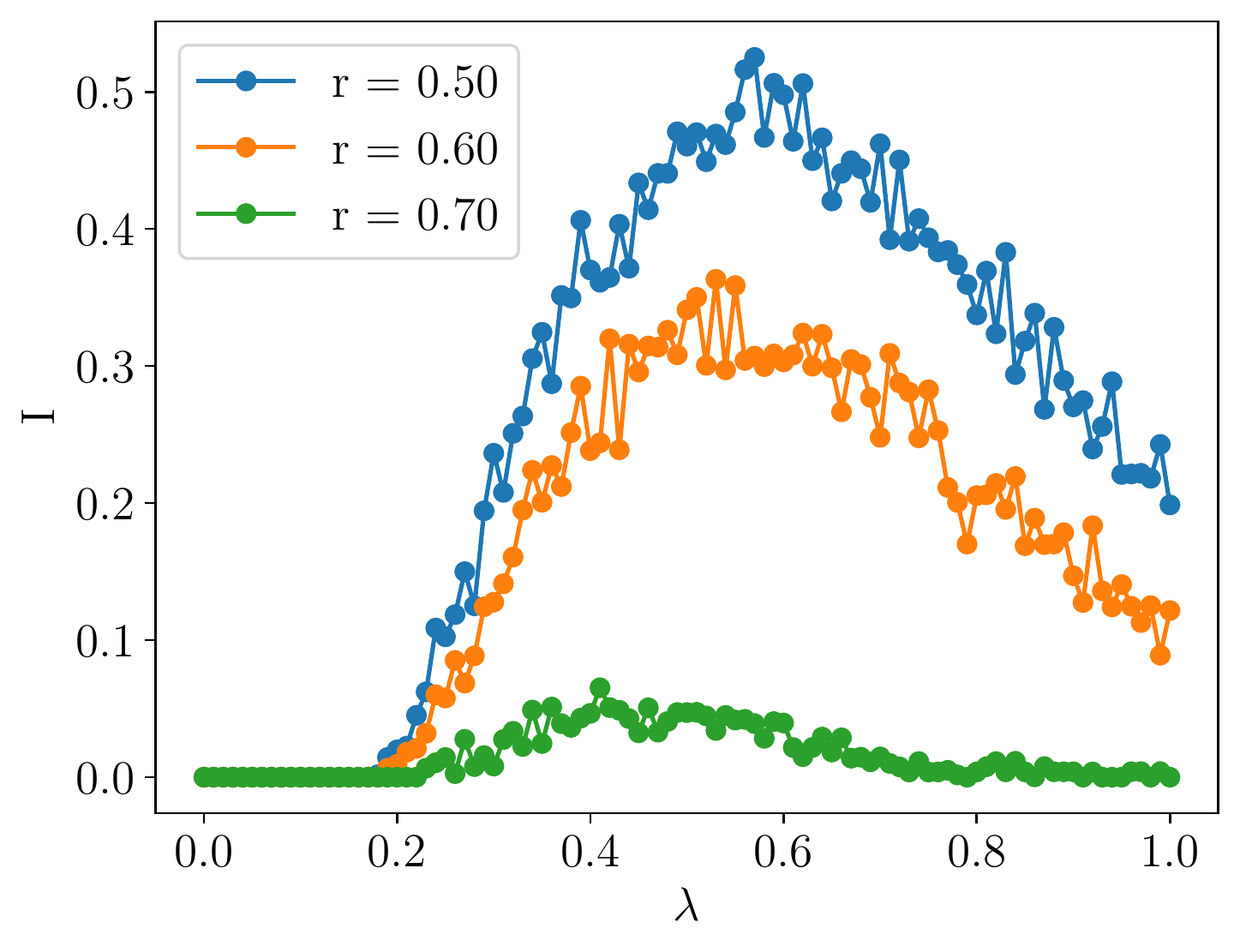}
    }
    %\subfloat[$w=0.4$]{
    %    \includegraphics[width=0.32\textwidth]{figs/ER_IxL_ac_w_0.400000_D_0.400000.pdf}
    %}
    \caption{Infected fraction of the population in red and fraction of samples that ended in the absorbing state in blue in (a), average infection rate over samples in (b) versus the infection rate. We use $D=0.4$ and $w=0.2$ in both and $r=0.5$ in (a). Here we observe a distinctly non-monotonic behavior in the average.}
    \label{IxLnonlin}
\end{figure}

Another non-monotonic effect can be observed in \cref{IxLnonlin}, where we can see that increasing the infection rate for certain parameters only increases the fraction of infected individuals in the steady state up to a certain point. Further increases in the infection rate, however, lead to a reduction. This effect has been previously observed in a similar model \cite{2018piresOC} and is related to an increase in risk perception as the disease spreads faster.

\FloatBarrier

\section{Final Remarks}

In conclusion, our study on the interplay between opinion dynamics, vaccination, and epidemic spreading in a non-static network provides important theoretical and practical implications for public health policies.

From the perspective of non-equilibrium dynamics, our Monte Carlo simulations reveal a first-order phase transition with metastable states, indicating the existence of multiple stationary states. This result is significant because it demonstrates that small changes in network parameters can have a significant impact on the dynamics of the epidemic. This finding introduces a new mechanism for the emergence of metastability, which aligns with previous related literature~\cite{oestereich2020hysteresis,chen2019imperfect,1976elster}.

Our results demonstrate that the network structure evolves to favor homophily, ultimately leading to the complete fragmentation of the network in certain cases. This finding emphasizes the importance of understanding multi-coupled dynamics related to the spread of infectious diseases.

Furthermore, our study reveals an intriguing phenomenon where an increase in the rewiring probability can increase the short-term epidemic peak due to the breakup of the network into smaller disconnected networks. However, in the long term, an increase in the rewiring probability leads to a decrease in the infection rate. These findings suggest that public health campaigns aimed at controlling epidemics through vaccination programs should consider the optimal level of rewiring in dynamic networks. Moreover, this finding adds to the list of counter-intuitive phenomena that can arise in vaccination dynamics~\cite{chen2019imperfect,zhang2012rational,wu2013peer,zhang2013braess,zhang2014effects,ichinose2017positive,steinegger2018interplay}.

Overall, our study provides valuable insights into the complex dynamics of infectious disease spread in dynamic networks with vaccination and opinion dynamics. The theoretical and practical relevance of this work can contribute to the design of effective public health policies for controlling epidemics in real-world scenarios.

\vspace{1cm}

\end{document}